\documentclass[]{aa} 
\pdfoutput=1 
\usepackage{graphicx}
\usepackage[varg]{txfonts}
\usepackage{color}

\usepackage{subfig}
\usepackage[percent]{overpic}
\usepackage[pdftex, pdfauthor={Maxime Trebitsch}, pdftitle={Polarization around LABs}, colorlinks,citecolor=blue]{hyperref}
\bibpunct{(}{)}{;}{a}{}{,}

\defcitealias{Rosdahl2012}{RB12}
\defcitealias{Hayes2011}{H11}
\defcitealias{Dijkstra2008}{DL08}

\def\RB{\citetalias{Rosdahl2012}\xspace}
\def\Hayes{\citetalias{Hayes2011}\xspace}
\def\DL{\citetalias{Dijkstra2008}\xspace}

\def\mclya{\textsc{MCLya}\xspace}
\def\ramses{\textsc{Ramses}\xspace}
\def\ramsesrt{\textsc{Ramses-RT}\xspace}
\def\lya{\ifmmode {\mathrm{Ly}\alpha} \else Ly$\alpha$\xspace \fi}
\def\Msun{\ensuremath{\mathrm{M}_\sun}\xspace}
\def\Rvir{\ensuremath{R_{\rm vir}}\xspace}
\def\hi{\ion{H}{i}\xspace}
\def\hii{\ion{H}{ii}\xspace}

\begin{document}

   \title{Lyman-$\alpha$ blobs: polarization arising from cold accretion}

%   \subtitle{}

   \author{Maxime Trebitsch
          \inst{1}
          \fnmsep\thanks{\email{maxime.trebitsch@ens-lyon.org}}
          \and
          Anne Verhamme\inst{2,1}
          \and
          J\'er\'emy Blaizot\inst{1}
          \and
          Joakim Rosdahl\inst{3}
          }

   \institute{Univ Lyon, Univ Lyon1, Ens de Lyon, CNRS, Centre de Recherche Astrophysique de Lyon UMR5574, F-69230, Saint-Genis-Laval, France
     \and
     Observatoire de Gen\`eve, Universit\'e de Gen\`eve, 51 Ch. des Maillettes, CH-1290 Sauverny, Switzerland
     \and
     Leiden Observatory, Leiden University, P.O. Box 9513, 2300 RA Leiden, the Netherlands
   }

   \date{Received ?? / Accepted ??}

% \abstract{}{}{}{}{} 
% 5 {} token are mandatory
 
  \abstract
  {
    % Context
    Lyman-$\alpha$ nebulae are usually found in massive environments at high redshift ($z \gtrsim 2$). The origin of their Lyman-$\alpha$ (\lya) emission remains debated. Recent polarimetric observations showed that at least some \lya sources are polarized. This is often interpreted as a proof that the photons are centrally produced, and opposed to the scenario in which the \lya emission is the cooling radiation emitted by gas heated during the accretion onto the halo.
    % Aims
    We suggest that this scenario is not incompatible with the polarimetric observations.
    % Methods
    In order to test this idea, we post-process a radiative hydrodynamics simulation of a blob with the \textsc{MCLya} Monte Carlo transfer code. We compute radial profiles for the surface brightness and the degree of polarization and compare them to existing observations.
    % Results
    We find that both are consistent with a significant contribution of the extragalactic gas to the \lya emission. Most of the photons are centrally emitted and scattered inside the filament afterwards, producing the observed high level of polarization. We argue that the contribution of the extragalactic gas to the \lya emission does not prevent polarization to arise. On the contrary, we find that pure galactic emission causes the polarization profile to be too steep to be consistent with observations.
    % Conclusion
  }


   \authorrunning{M. Trebitsch et al.}
   \titlerunning{Polarization around \lya blobs}

   \maketitle
%
%________________________________________________________________

\section{Introduction}
\label{sec:intro}

Spatially extended high redshift ($z \gtrsim 2$) Lyman-$\alpha$ nebulae (HzLAN) were discovered more than twenty years ago by \citet[][]{Chambers1990}, and then regularly observed around powerful radio sources \citep[][]{Heckman1991, vanOjik1997, Villar-Martin2002, Reuland2003, Villar-Martin2007}. In the early 2000s, \citet[][but see also \citealp{Francis1996,Fynbo1999,Keel1999}]{Steidel2000} found similar objects at $z \simeq 3$ which were not associated to radio galaxies. Their physical properties are very similar to the HzLANs previously found, with sizes up to a few hundred kilo-parsecs, and \lya luminosities up to $10^{44}\,\mbox{erg}.\mbox{s}^{-1}$. A few hundred of these HzLANs (called ``Lyman-$\alpha$ Blobs'' or LABs) have been found at $z = 2 - 6.5$ by recent surveys \citep[][]{Palunas2004, Matsuda2004, Matsuda2006, Dey2005, Saito2006, Ouchi2009, Yang2009, Yang2010, Prescott2013}. They are usually associated with quasars \citep{Bunker2003, Basu-Zych2004, Weidinger2004, Christensen2006, Scarlata2009, Smith2009, Overzier2013}, Lyman-Break Galaxies \citep{Matsuda2004}, infrared or sub-millimetre sources \citep{Chapman2001, Dey2005, Geach2005, Geach2007, Geach2014, Matsuda2007}. For some of these LABs, no galactic counterpart has been found \citep[][see also the ``Blob 6'' of \citealp{Erb2011}]{Nilsson2006}. All these associations support the consensus that HzLANs live in massive haloes which are in the densest regions of the Universe \citep{Steidel2000, Matsuda2004, Matsuda2006, Prescott2008, Saito2015}.

These observations raise two fundamental questions : where do the vast quantities of emitting gas come from, and what sources of energy power the observed \lya{} emission ? It has become clear during the last decade that a significant fraction of the gas in massive haloes at high redshifts is cold \citep[see e.g.][]{Birnboim2003,Kerevs2005,DekelBirnboim06,OcvirkPichonTeyssier08,BrooksEtal09,Kerevs2009,vandeVoortEtal11,vandeVoortSchaye12}. Simulations suggest that this cold gas reservoir is a complex mixture, dominated in mass by primordial accretion streams and tidal streams from galaxy interactions, and it is probably this gas that we see shine in HzLANs. The second question remains largely open, however, and it is unclear which energy source triggers (or sustains) \lya{} emission in this gas. There are basically three scenarios. The \lya{} radiation may be due to rapid cooling following shock-heating of this gas by large galactic outflows \citep[e.g.][]{Taniguchi2000,Ohyama2003,Mori2004,Geach2005}. Alternately, the \lya{} radiation may be emitted by recombinations that follow photo-ionization from the intergalactic ultraviolet background \citep{Gould1996} or from local sources \citep{Haiman2001, Cantalupo2005, Cantalupo2014, Kollmeier2010}. Finally, the \lya{} radiation may trace the dissipation of gravitational energy through collisional excitations as gas falls towards galaxies \citep{Haiman2000a,Fardal2001, Furlanetto2005, Dijkstra2006, Dijkstra2009,Faucher-Giguere2010, Goerdt2010, Rosdahl2012}.
%{\bf A nice feature of this model is the decoupling between the origin of the \lya{} radiation in the observed blob and the associated source.}
We emphasize that the cold stream scenario has been suggested in response to the large variety of sources LABs are associated with. Indeed, it gives a unified mechanism to explain the \lya{} radiation for this variety of sources, without relying on the presence of e.g. an active galactic nucleus (AGN) associated with the blob \mbox{\citep[see][]{Dijkstra2009}}.
This latter scenario, often dubbed the ``cold stream scenario'' is the focus of the present paper.

Because simulations that describe any of these three likely contributions to the luminosity of HzLANs are so uncertain, one would like to find observables that would help separate them observationally. It has been shown that scattering may lead to a polarized \lya{} emission around high-redshift galaxies and collapsing haloes \citep{Lee1998,Rybicki1999,Dijkstra2008}, and the degree of polarization of HzLANs may indeed help us disentangle the emission processes. In particular, it was argued by \citet{Dijkstra2009} that emission from cold accretion streams would not produce a polarized signal structured at the scale of the blob, because of the small volume filling of such streams. The first positive observation by \citet[hereafter \citetalias{Hayes2011}]{Hayes2011} of linear polarization forming a large-scale ring pattern around the \lya{} peak of LAB1 \citep{Steidel2000} was interpreted indeed as a strong blow against the cold stream scenario. \citet{Humphrey2013} further found the same level of polarization around the radio galaxy TXS 0211-122. Earlier observations from \citet{Prescott2011} showed no evidence for polarization around the HzLAN ``LABd05'', but \Hayes argued that this is due to a too low signal-to-noise ratio.

In the present paper, we revisit the question of the polarization of HzLANs from a theoretical perspective. We wish to test whether the results of e.g. \citet[hereafter \citetalias{Dijkstra2008}]{Dijkstra2008}, based on idealised geometrical configurations, hold when the full complexity of the cosmological context is taken into account. This will allow us to provide an alternative key to interpret polarization constraints, taking into account this complexity. We do this by extending the work of \citet[hereafter \citetalias{Rosdahl2012}]{Rosdahl2012} to assess whether their state-of-the-art simulation of a typical LAB (their halo H2, of mass $\sim 10^{12}\,\mbox{M}_\odot$ at $z=3$) is compatible with the observations of \Hayes{}. We show that it is indeed, and hence use it to discuss the composite origin of the polarization feature \Hayes observe. 

In Sec. \ref{sec:methodo}, we present the details of the simulation of \RB{} that we use, and we describe our new version of the Monte Carlo radiative transfer code \mclya \citep{Verhamme2006} which now tracks polarization of \lya photons. We then discuss in detail how we sample the emission from gas and stars in the simulation, and how we build polarization maps from the results of \mclya. In Sec.~\ref{sec:results}, we present our results and compare them to observations of \Hayes. We then discuss the origin of the polarization signal in our simulated nebula. Finally, we conclude in Sec.~\ref{sec:conclusions}. 

%__________________________________________________________________

\section{Methodology}
\label{sec:methodo}

\subsection{Description of the RHD simulation}
\label{sec:simu}
This work is based on the H2 simulation taken from \RB, which is our best model for a typical giant LAB at redshift 3. This simulation describes a halo of $\sim 10^{12} M_\odot$ at $z\sim 3$, which is a group of galaxies penetrated by cold accretion streams. We refer the reader to \RB for a full description of the numerical details and a discussion of the physical processes at play in that halo. 

In short, this simulation was performed with \ramsesrt \citep{Rosdahl2013}, a modified version of the adaptive mesh refinement (AMR) code \ramses \citep{Teyssier2002}, which couples radiative transfer of ultraviolet photons to the hydrodynamics. This radiation hydrodynamics (RHD) method allows to precisely follow the ionisation and thermal state of the intergalactic and circumgalactic media (IGM, CGM), accounting for self-shielding of the gas against the UV background, and thus to accurately compute the \lya emissivity of the gas \citep[see discussions in][]{Faucher-Giguere2010,Rosdahl2012}. We used a zoom  technique to achieve a maximal resolution of $434\, \mbox{pc}$ at $z = 3$, with a dark matter mass resolution of $1.1 \times 10^7\,\Msun$. Note that the refinement criteria we chose are such that the highest resolution is not only reached in the high-density ISM, but also along the cold streams. 

For the analysis below, we define the star-forming interstellar medium (ISM) as the gas denser than $n_{\mathrm{H}} \geq 0.76\,\mbox{cm}^{-3}$, the circum-galactic medium (CGM) as the gas at density $0.23 \,\mbox{cm}^{-3} \leq n_{\mathrm{H}} < 0.76\,\mbox{cm}^{-3}$, and the accretion streams as $0.015 \,\mbox{cm}^{-3} \leq n_{\mathrm{H}} < 0.23\,\mbox{cm}^{-3}$. We refer to the gas with lower density as diffuse gas. Note that these selections, although they rely on density only (and not temperature), do nicely pick out cold streams. This is likely in part due to the relative simplicity of the CGM in our simulation which do not include feedback from supernovae. 

\subsection{Polarized \lya radiative transfer: \mclya}
\label{sec:mclya}
The simulated halo H2 described in Sec.~\ref{sec:simu} is post-processed using an improved version of the Monte Carlo \lya transfer code, \mclya \citep{Verhamme2006}. Most of the improvements are discussed in \citet{Verhamme2012}: \mclya now makes use of the AMR structure of \ramses and includes more detailed physics for the \lya line. The new version of the code we use here introduces the ability to propagate photons emitted by the gas (see Sec.~\ref{sec:sources}), and most importantly to track the polarization state of Monte Carlo photons. 

As pointed out by \DL, the precise atomic level involved in the scattering of a \lya photon is strongly correlated with the scattering phase function: the $1S_{1/2} \rightarrow 2P_{1/2} \rightarrow 1S_{1/2}$ (K transition) scattering sequence is described by an isotropic phase function, losing therefore any polarization information. On the contrary, the $1S_{1/2} \rightarrow 2P_{3/2} \rightarrow 1S_{1/2}$ (H transition) sequence keeps a memory of the pre-scattering state of the photon. \citet{Hamilton1947} showed that, when they happen close enough to line centre (i.e. in the core), H transitions are well described by a superposition of an isotropic phase function and a Rayleigh phase function, with equal weights. \citet{Stenflo1980} later showed that for a scattering event outside of the \lya line centre (i.e. in the wings), the two transitions H and K interfere, and the event can instead be described by a single Rayleigh phase function.%\footnote{\label{fn:HK}The separation between the H and K transitions is very small: $\lambda_K~-~\lambda_H \lesssim 0.01\,\AA$. As soon as $\lambda - \lambda_{\lya} \gtrsim 0.02\,\AA$, the result interferences between the two transitions are very well described by a Rayleigh transition.}.
A convenient way to express the frequency is through its Doppler shift with respect to the \lya line centre, $x = (\nu-\nu_{\lya})/\Delta\nu_D$, where $\Delta\nu_D = (v_{th}+v_{turb}) \nu_\lya / c$. In this formula, $\nu_{\lya} = 2.466\, \mbox{Hz}$ ($\lambda_\lya = 1215.668\, \AA$) is the \lya line frequency, $v_{th}$ is the thermal velocity of hydrogen atoms, $v_{turb}$ is a turbulent velocity, describing the small scale turbulence of the gas, and $c$ is the speed of light.
\citet[appendix A2]{Dijkstra2008} show that in a Monte Carlo simulation, if we compute the photon frequency in the frame of the atom involved in the scattering event, we can take $x_{crit} \simeq 0.2$ to separate these two regimes of the H transition (core and wings). We follow their recommendation in the present paper, as we discuss later in this section.

There are mainly two approaches to describe the polarization state of light in a Monte Carlo framework. One possibility would be to consider groups of photons and compute the Stokes vector after each interaction as the result of a multiplication with a scattering matrix \citep{Code1995,Whitney2011}.
The other possibility is to use the technique described by \citet{Rybicki1999}, which is the one we implemented in this paper.
In this formalism, each Monte Carlo photon has a 100\% linear polarization given by a unit vector $\vec{e}$ orthogonal to the propagation direction $\vec{n}$ of the photon: $\vec{e}\cdot\vec{n} = 0$. The observed Stokes parameters will arise from the sum of multiple, independent MC photons.
Their initial scheme is only valid for a Rayleigh scattering event, but can be easily modified to take resonant scattering into account, since resonant scattering is described by a superposition of Rayleigh and isotropic scattering.

For scattering events in the line core, the probability of a K transition is $1/3$, and $2/3$ for an H transition. The H transition is described by 50\% of Rayleigh scattering and 50\% of isotropic scattering, and a K transition always corresponds to an isotropic scattering. This implies that for core photons (having $|x| < 0.2$ in the atom's frame) $2/3$ of the scattering events are actually isotropic (i.e. lose polarization), while $1/3$ are Rayleigh scatterings. For wing photons ($|x| > 0.2$), all scatterings are Rayleigh scatterings.

For an isotropic scattering event, the direction of propagation after scattering $\vec{n'}$ and the new direction of polarization $\vec{e'}$ are both randomly generated: $\vec{n'}$ is uniformly drawn on a sphere, and $\vec{e'}$ is uniformly drawn on a unit circle in a plane orthogonal to $\vec{n'}$.

For a Rayleigh scattering event, it can be shown \citep[see for instance][Appendix A3; also \citealt{Rybicki1999}]{Dijkstra2008} that the phase function can be simulated using a rejection technique: a random direction $\vec{n'}$ and a random number $R \in [0,1[$ are drawn, and the new direction is accepted if $R < 1 - \left(\vec{e}\cdot\vec{n'}\right)^2$. Otherwise, a new direction and a new number are drawn again. The new polarization vector $\vec{e'}$ is given by the projection of the previous polarization vector $\vec{e}$ on the plane normal to $\vec{n'}$:
\begin{equation}
  \label{eq:newpolarization}
  \vec{e'} = \frac{\vec{e} - \left(\vec{e}\cdot\vec{n'}\right)\vec{n'}}{\|\vec{e} - \left(\vec{e}\cdot\vec{n'}\right)\vec{n'}\|}.
\end{equation}

\subsection{\lya sources}
\label{sec:sources}

One of the motivation of this work is to understand if polarimetric observation can be a tool to elucidate the origin of the \lya emission of blobs (extended or centrally concentrated). 
We decompose the total \lya emission in two components: the {\it extragalactic} part is emitted by gas at densities $n_{\mathrm{H}} \leq 0.76\,\mbox{cm}^{-3}$, and is thus composed of CGM, cold streams and more diffuse gas, and the {\it galactic} part corresponds to the photons emitted by galaxies, i.e. form material at densities $n_{\mathrm{H}} \geq 0.76\,\mbox{cm}^{-3}$.

In our transfer code, a Monte Carlo photon is defined by a few quantities: position, propagation and polarization directions, luminosity and frequency. The initial propagation direction of a photon is randomly drawn on a sphere. This defines the initial polarization plane, in which lies the polarization direction (which is randomly drawn on a circle). The initial positions, luminosity and frequency of a photon are source-dependent. For the extragalactic emission, the photons will be emitted directly from the simulation cells, and the luminosity and frequency will be computed from the gas properties (see Sec.~\ref{sec:gaslya}). For the galactic emission, we use the star particles from the simulation as a proxy for the \lya sources. Their luminosities and frequencies are computed as explained in Sec.~\ref{sec:modelism}.

\subsubsection{Lyman-$\alpha$ emission processes}
\label{sec:lyaemission}

\lya emission is generated by two channels: collisional excitation of an hydrogen atom, and recombination of a free electron on a \hii ion.

The collisional mechanism is the following: a free electron excites an \hi atom, which can relax to its ground state. During its radiative cascade, a $2P \rightarrow 1S$ transition may occur, causing the emission of a \lya photon. We approximate the collisional emissivity with
\begin{equation}
  \label{eq:lyacollisional}
  \varepsilon_{\mathrm{coll}} = C_{\lya}(T)\, n_{\mathrm{e}}\, n_{\hi}\, \epsilon_{\lya},
\end{equation}
where $n_{\mathrm{e}}$ and $n_\hi$ are the electron and \hi number densities, $\epsilon_\lya = 10.2\mbox{eV}$ is the \lya photon energy, and $C_\lya(T)$ is the rate of collisionally induced $1S\rightarrow 2P$ transitions. We use the expression given by \citet{Goerdt2010} for $C_\lya(T)$, fitting the results from \citet{Callaway1987}.

The recombination process occurs when a free electron recombines with a proton to give an excited hydrogen atom. This atom may cascade down to the $2P$ level from its excited state, eventually relaxing to the ground state and producing a \lya photon. The \lya emissivity of the process is given by
\begin{equation}
  \label{eq:lyarecombination}
  \varepsilon_{\mathrm{rec}} = 0.68\, \alpha^{\mathrm{B}}_\hi (T) \, n_{\mathrm{e}} \, n_\hii \, \epsilon_\lya,
\end{equation}
where the $0.68$-factor is the average number of \lya photon produced per case B recombination for a typical gas temperature of $10^4\ \mbox{K}$ \citep{Osterbrock2006}, $n_\hii$ is the proton number density, and $\alpha^{\mathrm{B}}_\hi$ is the case B recombination rate taken from \citet{Hui1997}.

\subsubsection{Sampling the galactic emission}
\label{sec:modelism}
As stated in Sec.~\ref{sec:simu}, the simulation in \RB can only resolve physical processes at the scale of a few hundred parsecs.
This resolution is far from allowing us to resolve the interstellar medium structure of galaxies \citep[see][]{Verhamme2012}, and we thus have to use a model for the \lya luminosities and line profiles of our simulated galaxies. We use young star particles as a proxy for emission from \hii regions, and assign each particle younger than $10\, \mbox{Myr}$ a luminosity given by $(\text{particle mass}/\text{10 Myr})\ \times\ 1.1 \times 10^{42}\ \mbox{erg.s}^{-1}$ \citep{Kennicutt1998,Osterbrock2006}.  Guided by the results of \cite{Garel2012}, we assume a 5\% \lya escape fraction, typical of Lyman break galaxies. This implies that the galactic \lya luminosity in our simulation is about 30\% ($9 \times 10^{42}\ \mbox{erg.s}^{-1}$) of the total simulated LAB. To model the result of the complex \lya radiative transfer through the ISM, we use three different spectral shapes: a Gaussian plus continuum, with an equivalent width of 40 \AA, and two P-Cygni-like profiles, with the same equivalent width, but peaked at $250\,\mbox{km.s}^{-1}$ and $500\,\mbox{km.s}^{-1}$. We emulate the P-Cygni profiles with a ``Gaussian minus Gaussian'' function, plus a continuum. These line profiles describe the photons escaping from the galaxies of the simulation, which are then scattered through the CGM and more diffuse gas. To make this effective, we also render the ISM transparent to \lya photons. 

\begin{figure}
  \centering
  \resizebox{.9\hsize}{!}{\includegraphics{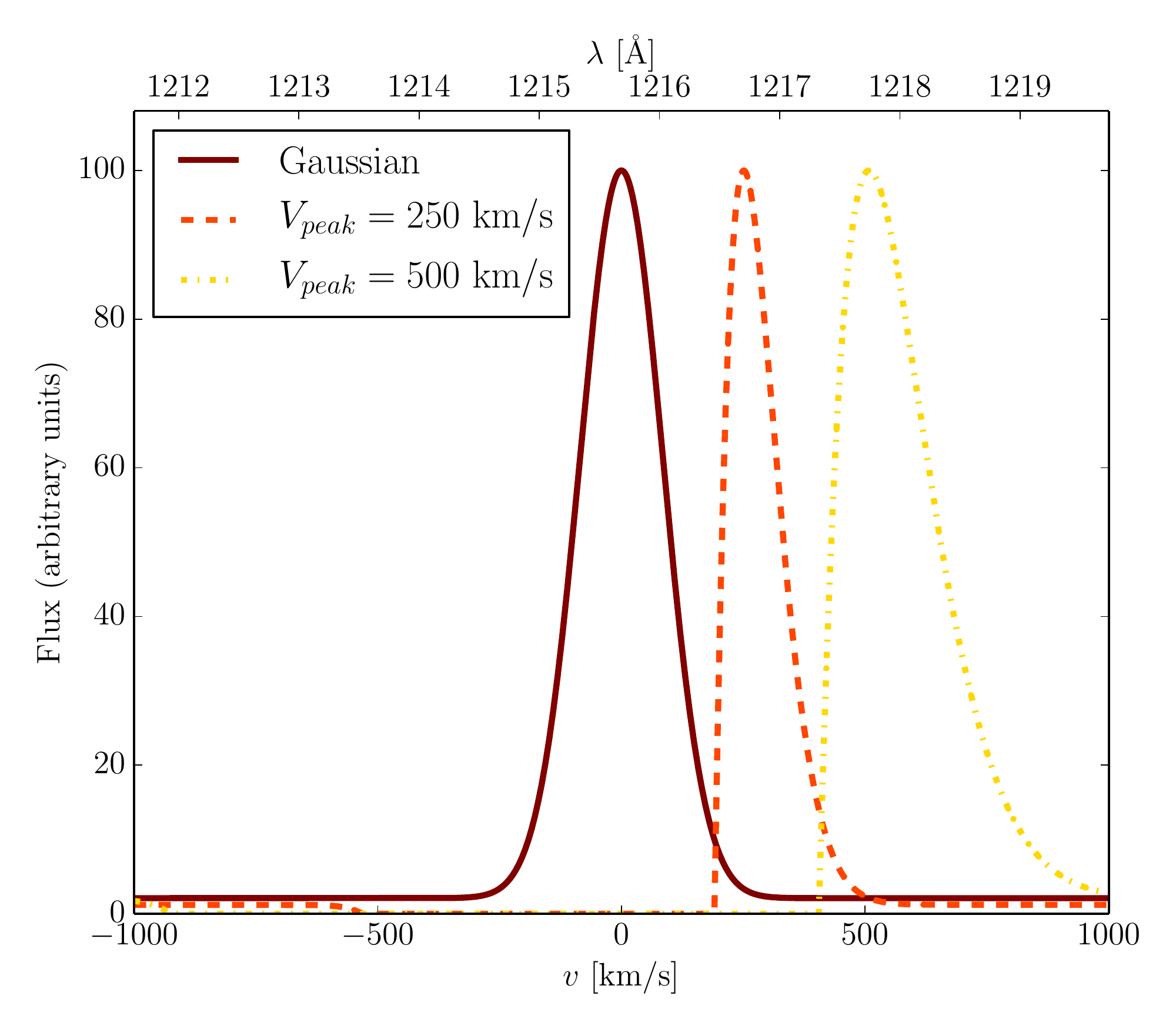}}
  \caption{Two different spectral shapes: a single Gaussian (solid), and a P-Cygni profile (dashed).}
  \label{fig:spectralshapes}
\end{figure}

The distribution of these star particles is presented in Fig.~\ref{fig:illustration_stars}, and the three line profiles are illustrated on Fig.~\ref{fig:spectralshapes}.

We found only little impact of the input spectral shape either on the surface brightness (SB) or on the polarization of the LAB. With the P-Cygni-like profiles, the degree of polarization tends to be slightly higher by a few percents, because much of the scattering gas is infalling. Hence, by using the Gaussian profile as our fiducial model for the \lya spectrum at the boundary of the ISM, we will get a lower limit of the estimated contribution of the galactic emission.

\subsubsection{Sampling the extragalactic gas emission}
\label{sec:gaslya}

In the blob simulation of \RB, the ionisation state of the gas, its temperature and the density are directly given by \ramsesrt, and the local emissivity of the gas is computed as $\varepsilon = \varepsilon_{\mathrm{coll}} + \varepsilon_{\mathrm{rec}}$. In this specific simulation, the extragalactic gas contributes to the total \lya luminosity by $L_{\mathrm{gas}} \simeq 2.11 \times 10^{43}\ \mbox{erg.s}^{-1}$. 36\% of luminosity of the extragalactic gas comes from the CGM, and more than 55\% comes from the streams.
As the luminosity of the gas varies by more than 12 orders of magnitude among $\sim 4 \times 10^6$ AMR cells, we cannot afford to sample the gas luminosity by sending from each cell a number of photons proportional to the cell luminosity. By sending at least 100 photons per cell, such a proportional sampling would require the prohibitive total of $10^{17}$ photons. We chose instead to send a fixed number of 150 photons from each of the $\sim 256\,500$ most luminous cells of the simulation. This restricts the range of luminosities to only three orders of magnitude. The average luminosity of the 100 faintest cells in our sample is approximately 2\,300 times lower that the average luminosity of the 100 brightest cells. Doing so, each photon will carry $\frac{1}{150}$ of its mother cell luminosity.
We evaluate the impact of our (under)sampling strategy of the simulation cells using a bootstrap method (see Appendix~\ref{sec:bootstrap} for details).

We fixed the limit of 256\,500 cells after ensuring that taking more gas into account would not noticeably affect our results. This (limited) set of cells still accounts for $\sim 97\%$ of the total blob luminosity ($L_{\mathrm{gas}} = 2.04 \times 10^{43}\ \mbox{erg.s}^{-1}$). Table~\ref{tab:budget} compares the luminosity budget for the whole halo and for the sampled cells.
As expected, the 256\,500 brightest cells that we cast photons from capture most of the luminosity of the CGM (99.9\%) and of the cold streams (97.4\%) but leaves out about a third of the luminosity of the very diffuse gas. What we miss from the very diffuse gas is a very small fraction ($\sim$ 1\%) of the total luminosity and has no impact on our results.

\begin{table}
\caption{Luminosity budget, in $10^{42}\ \mbox{erg.s}^{-1}$}
\label{tab:budget}
\centering
\begin{tabular}{c c c c}
\hline\hline
$L\, (10^{42}\ \mbox{erg.s}^{-1})$ & Total & Sampled  & Fraction\\
\hline
    CGM & $7.60$ & $7.59$ & 99.9\% \\
    Streams & $12.6$ & $12.3$ & 97.4\%\\
    Diffuse gas & $0.885$ & $0.575$ & 65.0\% \\
\hline
\end{tabular}
\end{table}

The last physical parameter to determine before casting a \lya photon is its exact wavelength. We draw the initial frequency of each Monte Carlo photon from a Gaussian distribution, centred on $\nu_\lya$ in the frame of the emitting cell. We set the width of this Gaussian line to be $\sigma_\lya = \nu_\lya \sqrt{v_{th}^2 + v_{turb}^2}/c$, where $v_{th}$ is the typical velocity of atoms due to thermal motions, and $v_{turb} = 10$ km$/$s describes sub-grid turbulence.  

In Fig.~\ref{fig:illustration_gas}, we illustrate the source distribution for the extragalactic emission.

\subsection{Mock observations}
\label{sec:observation}
In order to observe our simulated LAB, we collect the photons when they pass the virial radius. Photons exiting the halo are selected in a cone of 15\degr{} around the projection direction. We discuss the impact of the selection on the results in Appendix~\ref{sec:angle}. We then project these photons on a grid of 200 pixels on a side (equivalent to 0.125\arcsec{}).
We shall now describe how we build polarization maps from \mclya output.

We assume that each Monte Carlo photon is equivalent to a (polarized) beam of light, and that two independent photons are incoherent. Then, each pixel of our mock maps receives a mixture of independent, linearly polarized beams. The Stokes parameters are thus given by \citet[][eq.~(164),~§15]{Chandrasekhar1960}:
\begin{align}
  \label{eq:stokes}
  \begin{split}
    I &{}= \sum I^{(n)},\\
    Q &{}= \sum I^{(n)}\cos\left(2\chi_n\right),\\
    U &{}= \sum I^{(n)}\sin\left(2\chi_n\right),
  \end{split}
\end{align}
where $I^{(n)}$ defines the intensity of each beam, and $\chi_n$ is the polarization angle of each beam (with respect to a set of axes). Here, we have no $V$ Stokes parameter since we assumed a purely linear polarization for each Monte Carlo photon.

With Eq. \ref{eq:stokes}, we build the $I$, $Q$, and $U$ maps in a set of chosen directions from the output of \mclya, and we smooth them with a Gaussian of full width at half-maximum 1\arcsec{} to mimic a typical point spread function (PSF) in observations.

We extract the degree of polarization $\mathcal{P}$ and the angle of polarization $\chi$ in each pixel with
\begin{equation}
  \label{eq:poldegree}
  \mathcal{P} = \frac{\sqrt{Q^2+U^2}}{I}
\end{equation}
and
\begin{equation}
  \label{eq:polangle}
  \chi = \frac{1}{2}\mbox{arctan}\left(\frac{U}{Q}\right).
\end{equation}

Note that we compute the degree and angle of polarization in pixels with more than 5 MC photons after smoothing. This tends to overestimate the degree of polarization at high radius, but has no impact in the inner 40 kpc.

%__________________________________________________________________
\section{Results}
\label{sec:results}

\begin{figure*}
  \centering
  \subfloat{%
    \begin{overpic}[width=6.5cm]{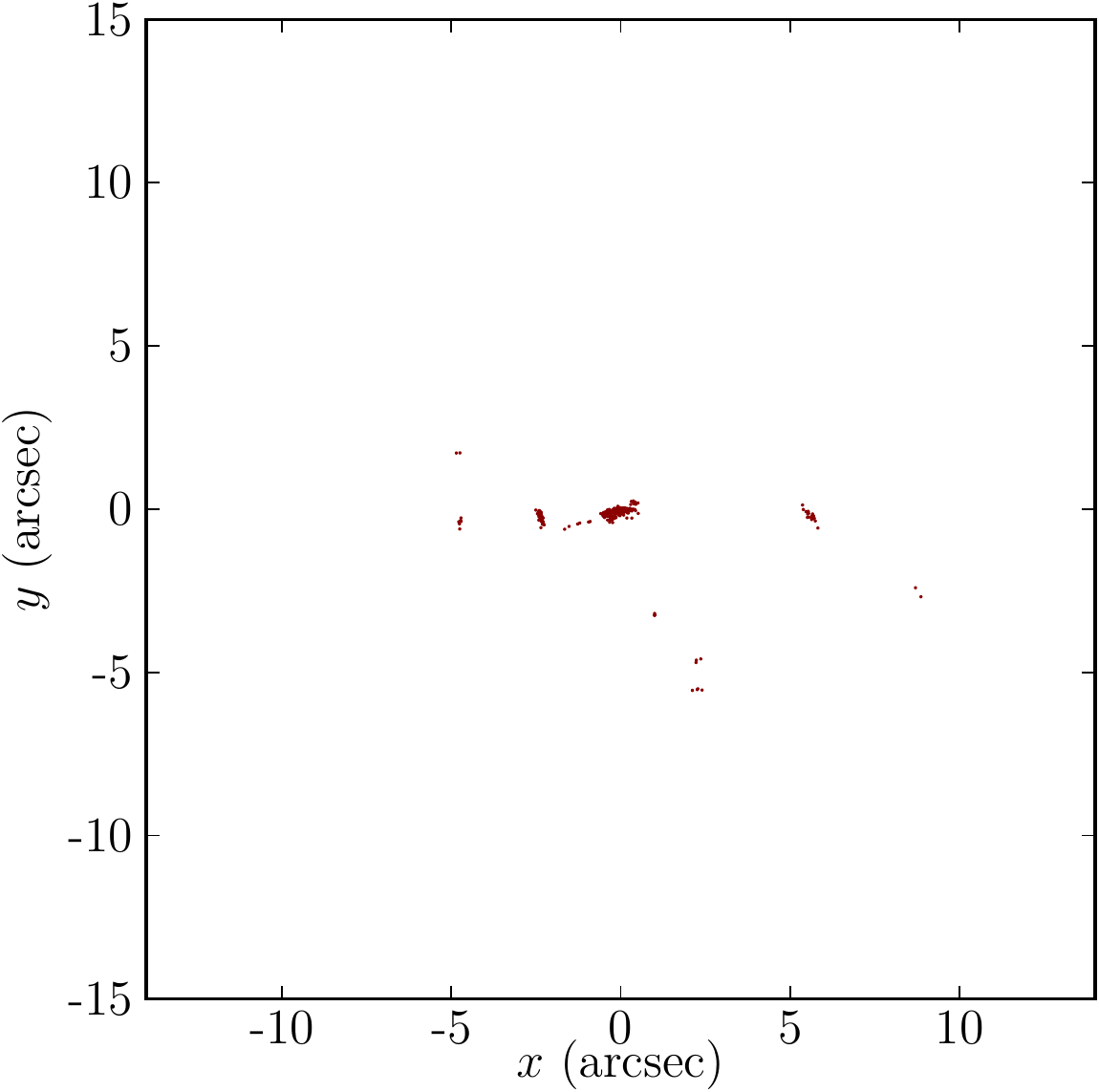}
        \put(80,15){\large\textbf{(a)}\protect\label{fig:illustration_stars}}
    \end{overpic}
  }
  \subfloat{%
    \begin{overpic}[width=6.5cm]{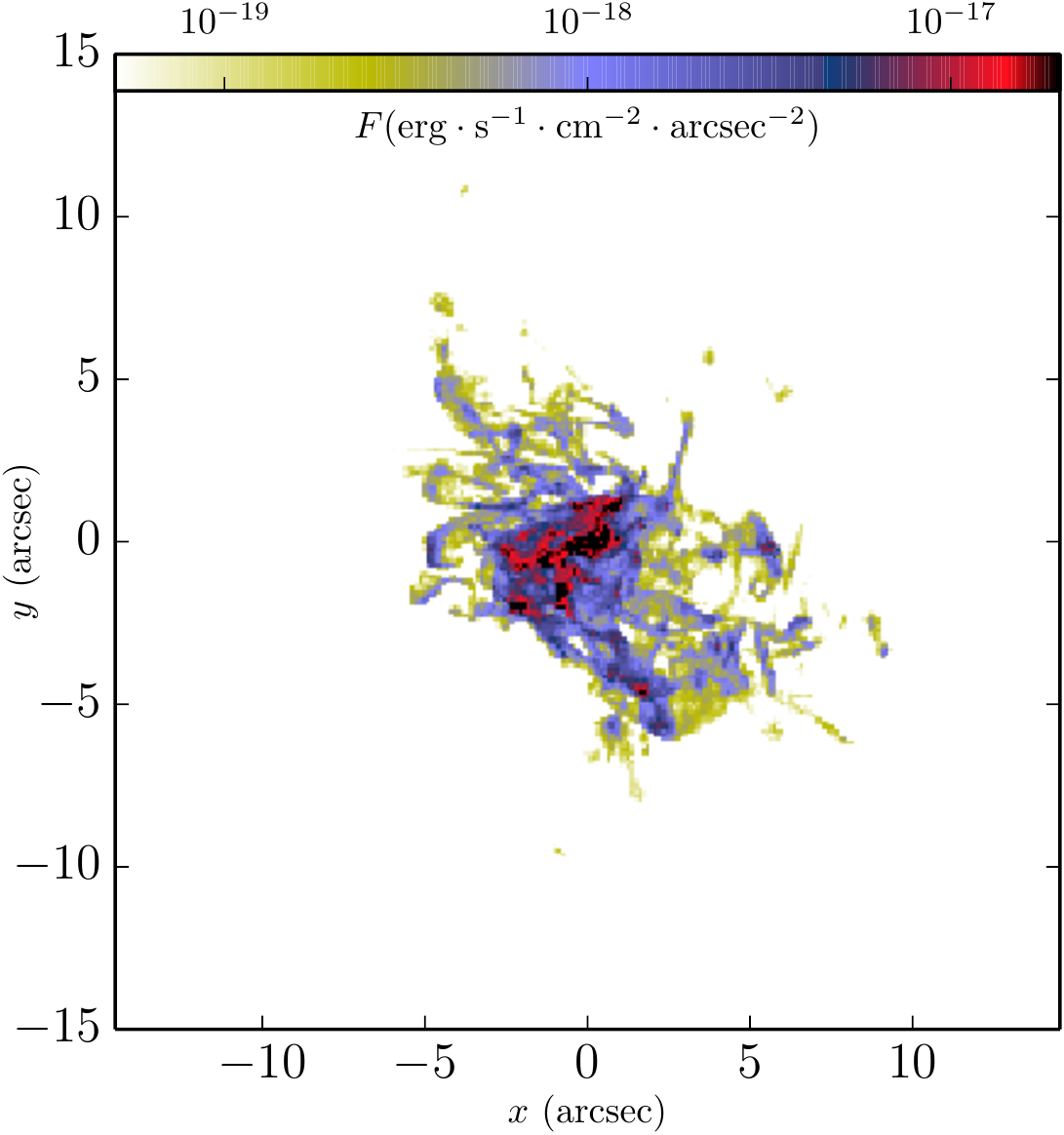}
        \put(80,15){\large\textbf{(b)}\protect\label{fig:illustration_gas}}
    \end{overpic}
  }\\
  \subfloat{%
    \begin{overpic}[width=6.5cm]{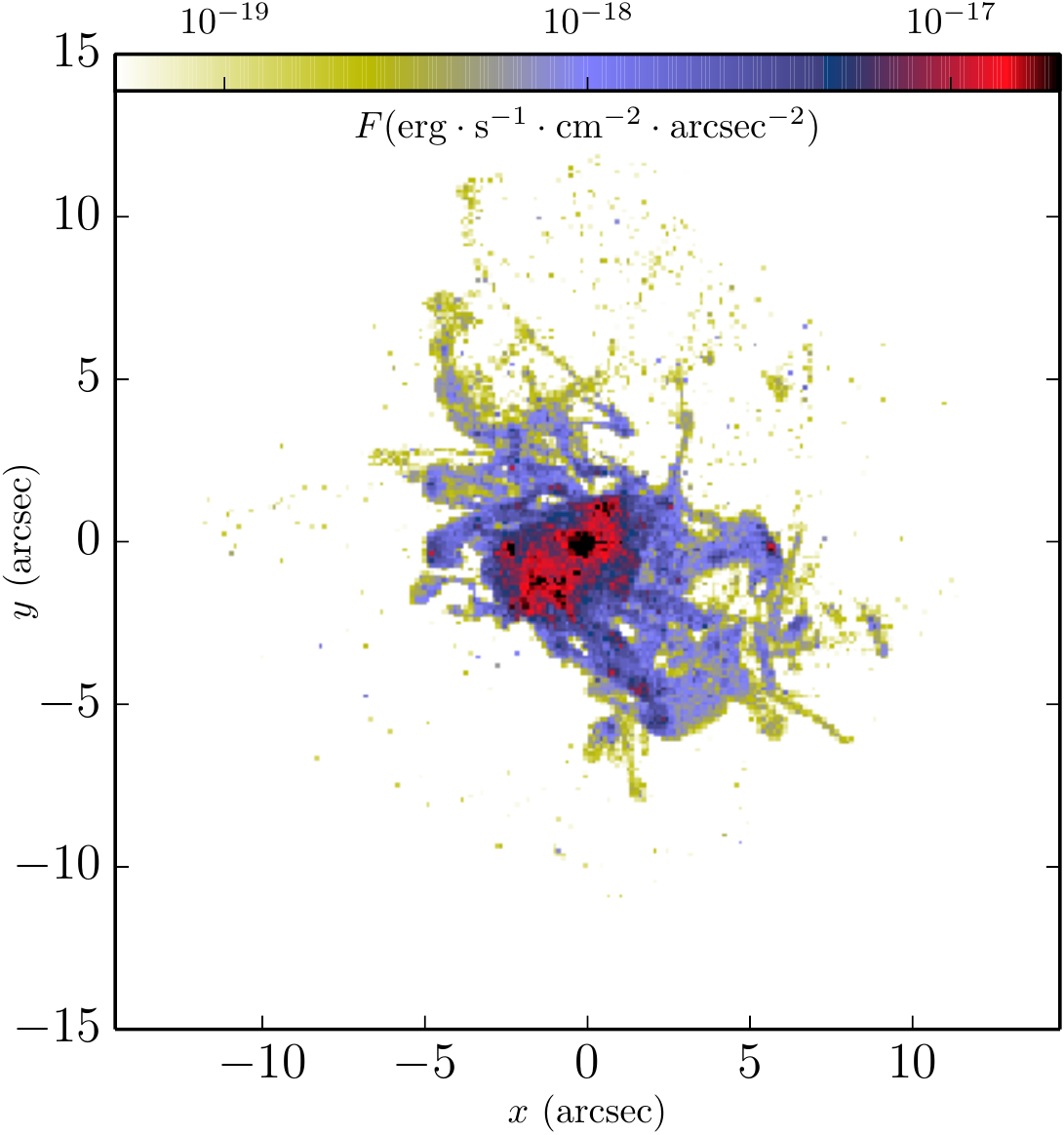}
        \put(80,15){\large\textbf{(c)}\protect\label{fig:illustration_nopsf}}
    \end{overpic}
  }
  \subfloat{%
    \begin{overpic}[width=6.5cm]{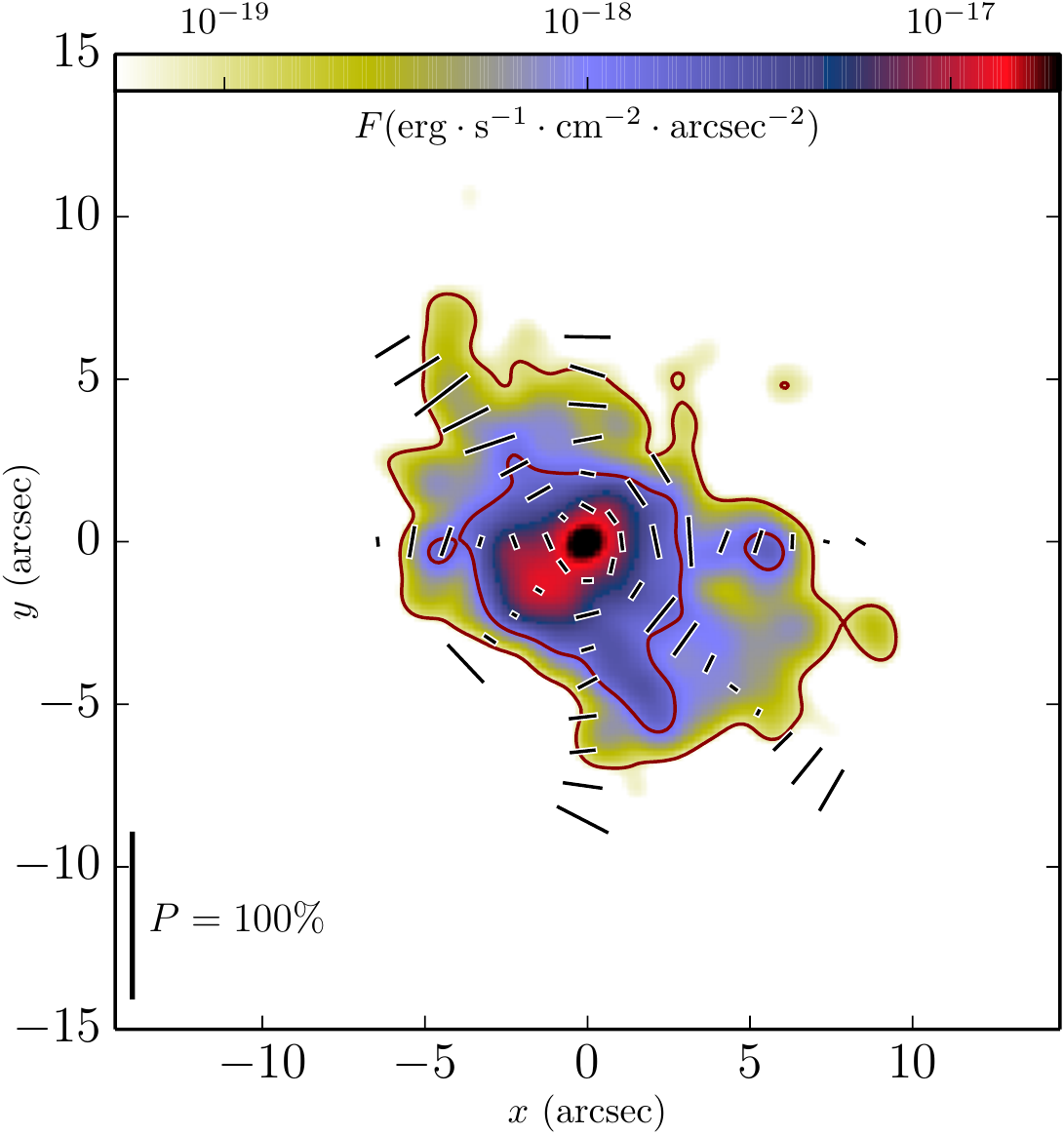}
        \put(80,15){\large\textbf{(d)}\protect\label{fig:illustration_results}}
    \end{overpic}
  }

  \caption{(a)~Positions of the young, massive stars from which we cast \lya photons that make the galactic contribution. (b)~SB map of the extragalactic emission region. (c)~SB map of the blob after transfer, with both galactic and extragalactic contribution to \lya emission. (d)~Mock observation of the halo with a seeing of $1\arcsec$. The dashes show polarization, and the contours mark $1.4\times 10^{-18}\,\mbox{erg} \ \mbox{s}^{-1} \ \mbox{cm}^{-2} \ \mbox{arcsec}^{-2}$ and $10^{-19}\,\mbox{erg} \ \mbox{s}^{-1} \ \mbox{cm}^{-2} \ \mbox{arcsec}^{-2}$.}
  \label{fig:illustration}
\end{figure*}

On Fig.~\ref{fig:illustration_results}, we show a mock image of our simulated blob. The inner iso-contours mark surface brightnesses of $1.4\times 10^{-18}\,\mbox{erg} \ \mbox{s}^{-1} \ \mbox{cm}^{-2} \ \mbox{arcsec}^{-2}$, which are typical of present observational limits. The outer contours at $10^{-19}\,\mbox{erg} \ \mbox{s}^{-1} \ \mbox{cm}^{-2} \ \mbox{arcsec}^{-2}$ show what we might see in deep VLT/MUSE observations. The bars show the polarization direction and amplitude (with a scaling indicated in the bottom-left corner of the plot) in different points chosen for illustration purposes\footnote{\label{fn:polimg}We only show the polarization signal in pixels having more than 10 Monte Carlo photons.}.

We now analyse our results and compare them to observations. Our workflow is the following. First, we produce mock observations along multiples lines of sight (LOS). Then, we compute SB profiles and polarization profiles for each LOS. Finally, we average the profiles over all LOS.

\subsection{Surface brightness profiles}
\label{sec:sbprofile}

\begin{figure}
  \centering
  \resizebox{.8\hsize}{!}{\includegraphics{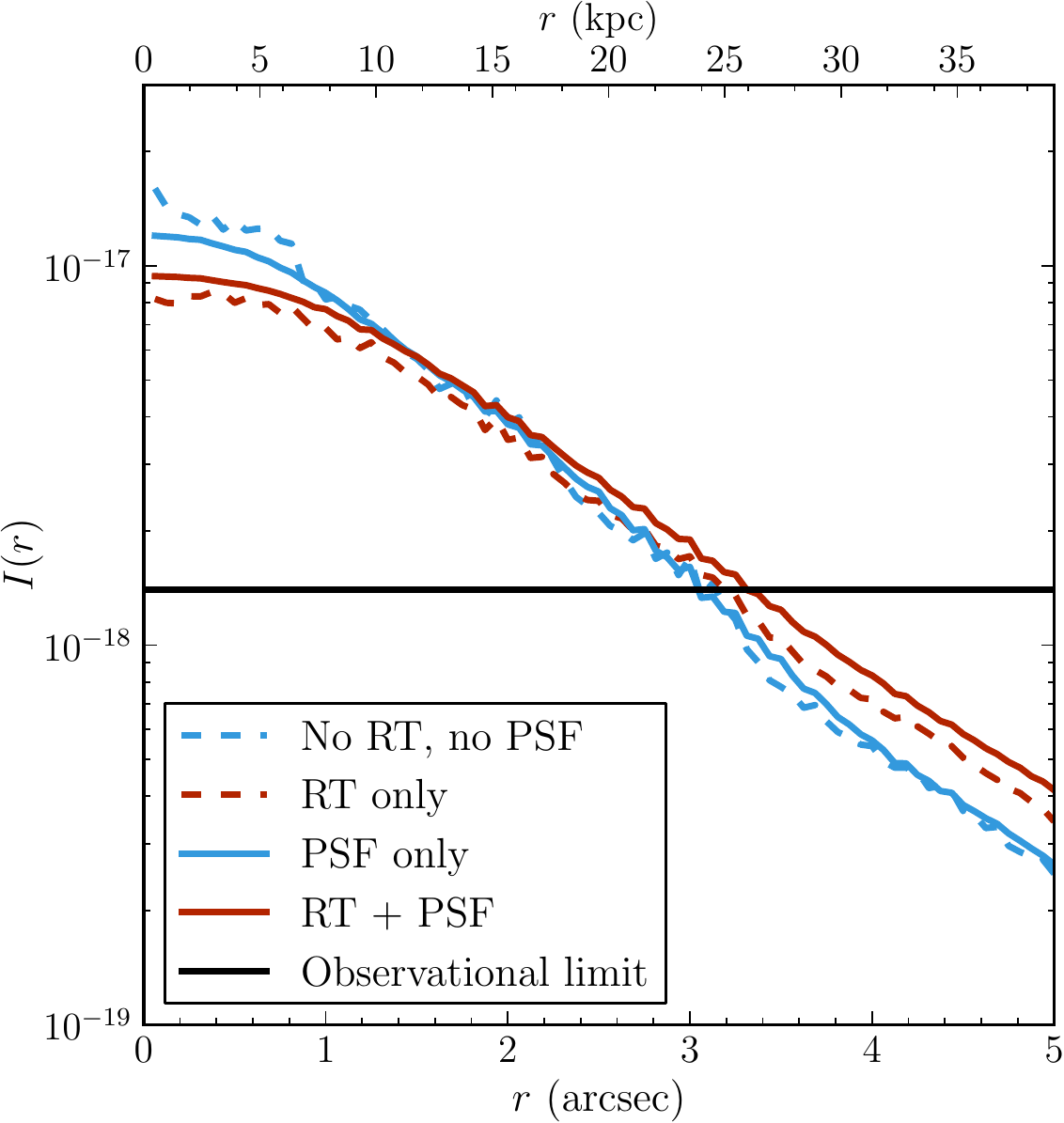}}
  \caption{Effect of scattering compared to that of a PSF on the observed SB profile for the extragalactic contribution to the luminosity.
The SB profiles before (after) transfer are shown in blue (red), the profiles with (without) PSF are represented as a solid (dashed) line. Note that only the emission from extra-galactic gas is taken into account here.}
  \label{fig:rt_effect}
\end{figure}

\RB argued that adding \lya scattering effects to their simulation would not change much the observed area of the blob. They also neglected the (galactic) contribution of star formation to the total \lya luminosity. With our simulation, we can compare the effect of scattering to that of a typical PSF on the observed SB profile for the extragalactic contribution to the luminosity of the LAB. On Fig.~\ref{fig:rt_effect}, we show the SB profile before and after transfer (in blue and in red, respectively), and before and after the convolution with a PSF (dashed line and solid line, respectively). We find that \lya{} scattering leads to a redistribution of light out to larger radii than a Gaussian PSF of 1 arcsec. This strongly impacts the inner ($r < 5$kpc) and outter ($r > 25$kpc) profile, as shown by the difference between the blue and red dashed curves. Coincidentally, however, at the level of $1.4\times 10^{-18}\,\mbox{erg} \ \mbox{s}^{-1} \ \mbox{cm}^{-2} \ \mbox{arcsec}^{-2}$, the effect of scattering is comparable to that of the PSF. At this surface brightness, we find that neglecting radiative transfer leads to an underestimate of the LAB's radius of only $\Delta R/R \simeq 8\%$, that is, a relative error on the blob area of $\Delta A/A \simeq 16\%$.

\begin{figure}
  \centering
  \resizebox{.8\hsize}{!}{\includegraphics{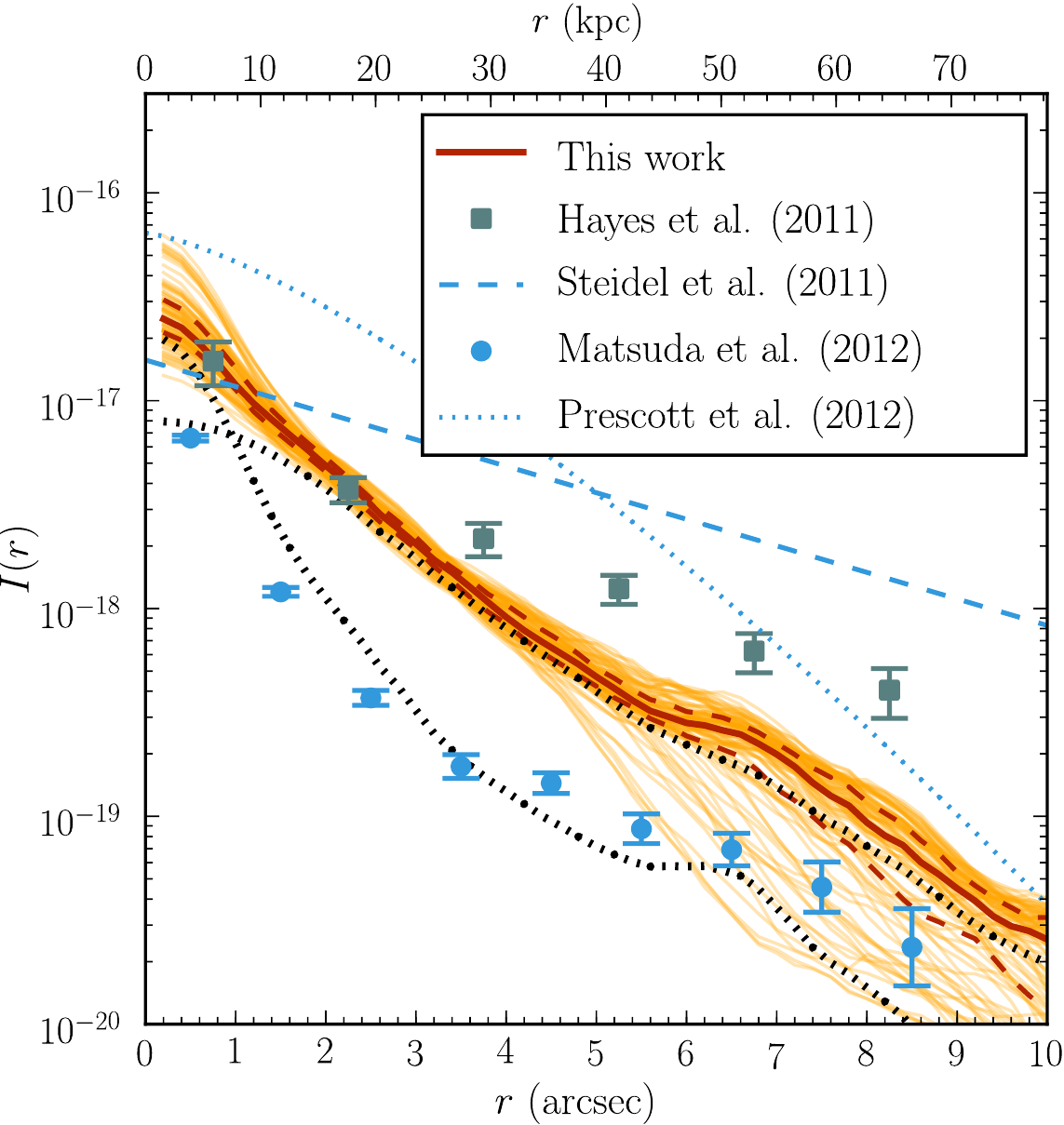}}
  \caption{Comparison of SB profiles. The red, solid line is the SB profile expected from the sum of both gas and galactic contributions; the red, dashed lines show the first and third quartiles. The thin, orange lines represents the profile for each LOS. The upper (lower) black, dotted line shows the the extra-galactic (galactic) contributions. Two observational data taken from the literature (see text box) are shown in blue, data points are \Hayes observations (teal) and \citet{Matsuda2012} stacked profile (blue).}
  \label{fig:sbgasstars}
\end{figure}

The observed surface brightness profiles provide a strong constraint on the properties of the HzLANs. On Fig.~\ref{fig:sbgasstars}, we show a comparison between the total surface brightness profile of our simulated blob (taking both galactic and extragalactic \lya emission into account, as discussed in Sec. \ref{sec:sources}) and a set of observational contraints. The thin, orange lines show the profiles of our simulated blob along each of the 100 lines of sight, and the thick, red, solid (resp. dashed) line shows median profile (first and third quartiles). We also plotted the galactic (lower dotted line) and extragalactic contributions (upper dotted line) to the luminosity. The galactic component dominates at the centre ($<1 \arcsec$) and is soon overtaken by extragalactic emission which represents about 90\% of the signal at all radii $>2$ arcsec. The blue dashed line the average profile of 11 LABs observed at $z = 2-3$ \citep{Steidel2011}, and the blue, dotted line is the fit given by \citet{Prescott2012} for LABd05, rescaled to $z = 3$ (however, LABd05 is 5 times brighter than our blob). We compare the results to the average surface brightness profile of a sample of 130 \lya emitters (LAE) in regions with a large LAE overdensity (blue circles) taken from \citet{Matsuda2012}. The teal squares show \Hayes observation of LAB1 \citep{Steidel2000}, rescaled so that its total luminosity is similar to H2.

From Fig.~\ref{fig:sbgasstars}, it seems that our profile agrees with \citet{Prescott2012}, and especially if we focus on the extragalactic contribution (upper black, dotted curve). We point out that this was already true for the profile with no scattering found by \RB (see their Fig. 13). The LAE used in the sample of \citet{Matsuda2012} are significantly smaller and fainter than our LAB, but they are remarkably similar to the galactic \lya emission of our simulation. Our result seems to be inconsistent with the profile from \citet{Steidel2011}. However, a good agreement was not expected: indeed, \RB argue that only their most massive halo H3 fit the results from \citet{Steidel2011}. Maybe more importantly since it is the only positive polarimetric observation, our results are in correct agreement with \Hayes data, though slightly steeper at large radii. 

\subsection{Polarization}
\label{sec:polarization}

For each plot in this section (Figs. \ref{fig:profiledir}, \ref{fig:anglehist}), the 100 LOS are depicted as thin, orange lines. The red, solid line represents the median profile. The interval between the two dashed, red lines contains 50\% of the LOS. With the bootstrap method described in App. \ref{sec:bootstrap}, we estimate the error due to our cell sampling strategy and show it as a red area around the median profile.

\subsubsection{Polarization profile}
\label{sec:polprofile}

\begin{figure*}
  \centering
  \subfloat{%
    \begin{overpic}[width=5.66cm]{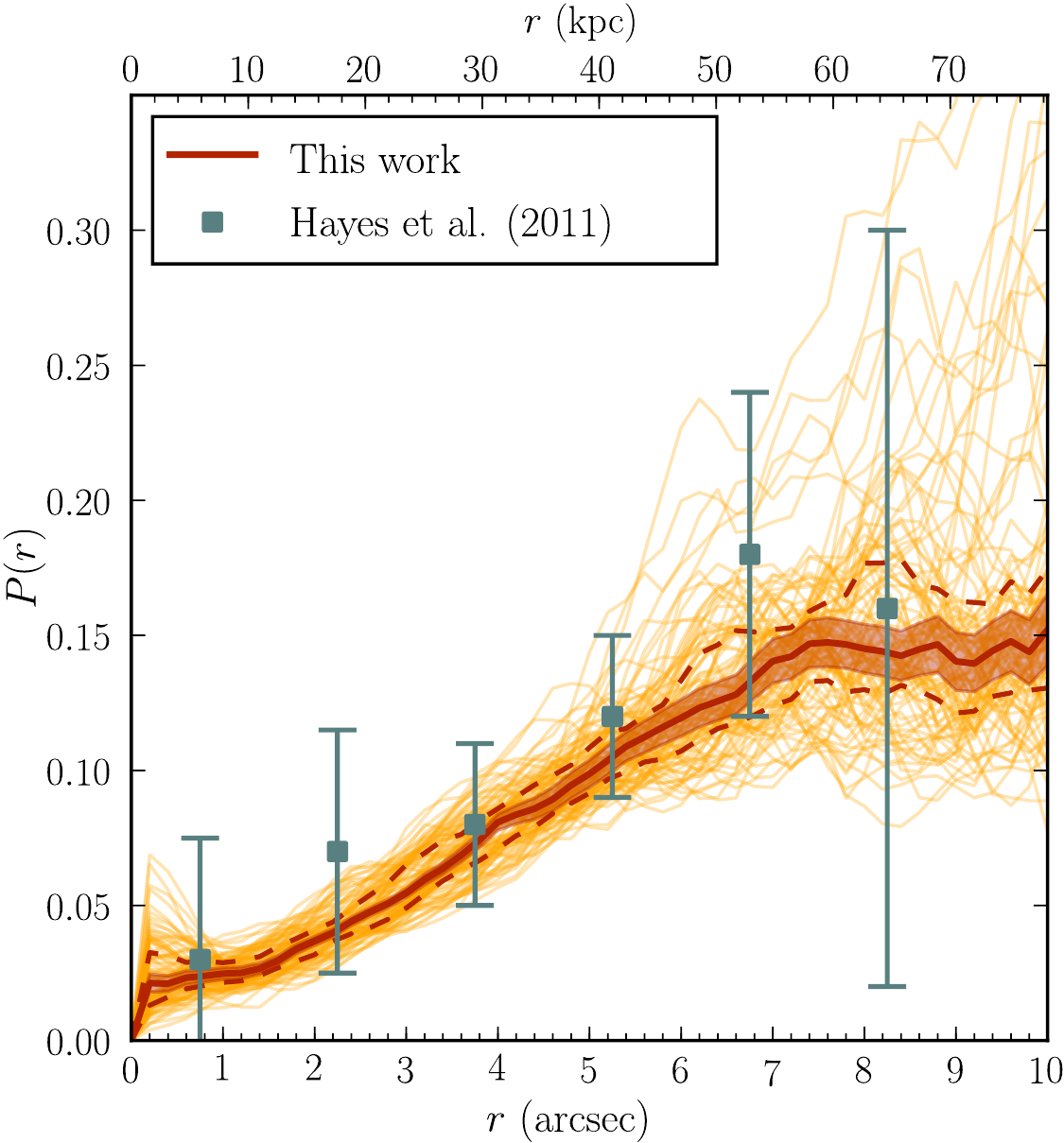}
      \put(80,15){\large\textbf{(a)}\protect\label{fig:profiledir_gas}}
    \end{overpic}
  }%
  \subfloat{%
    \begin{overpic}[width=5.66cm]{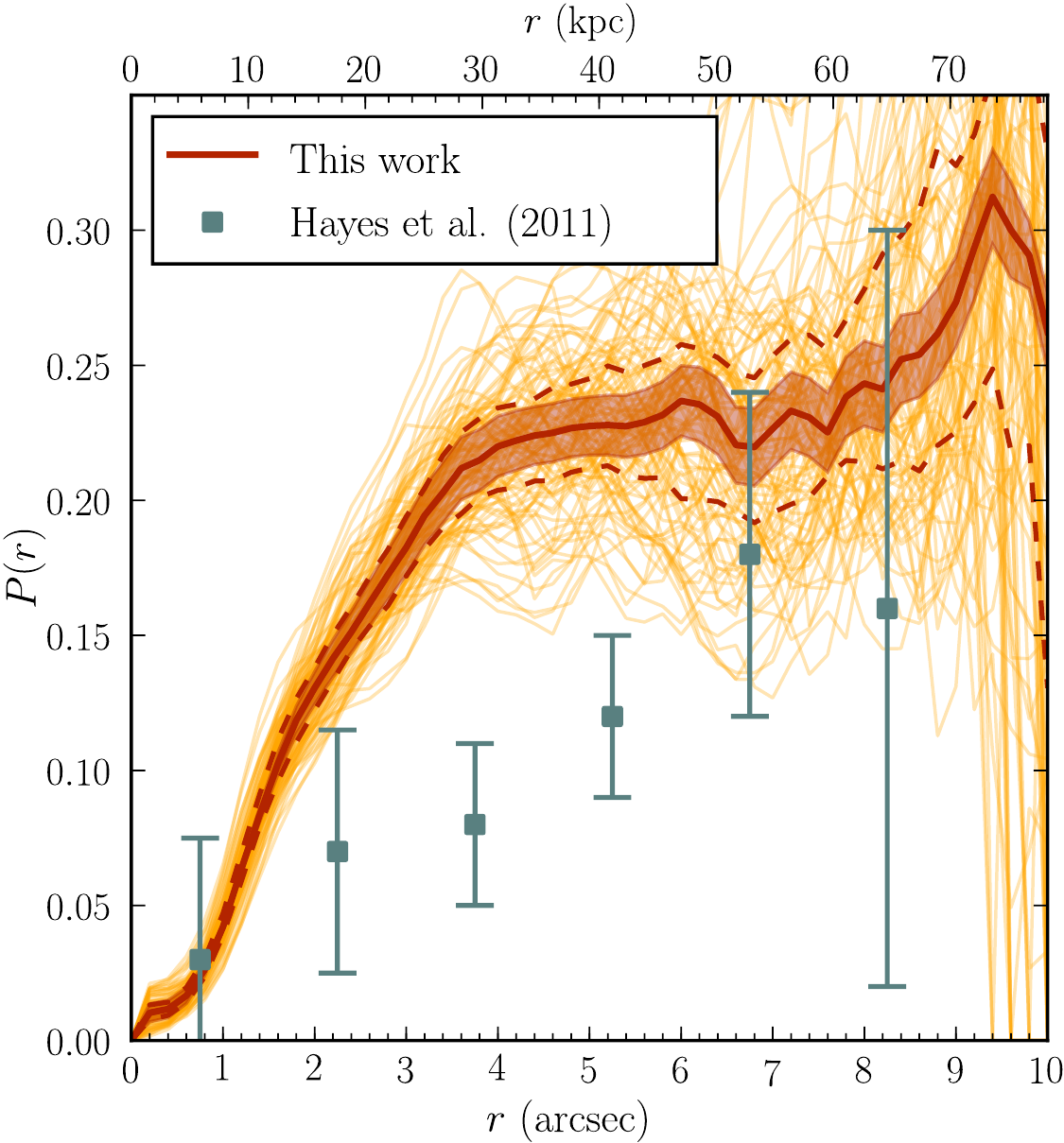}
      \put(80,15){\large\textbf{(b)}\protect\label{fig:profiledir_stars}}
    \end{overpic}
  }%
  \subfloat{%
    \begin{overpic}[width=5.66cm]{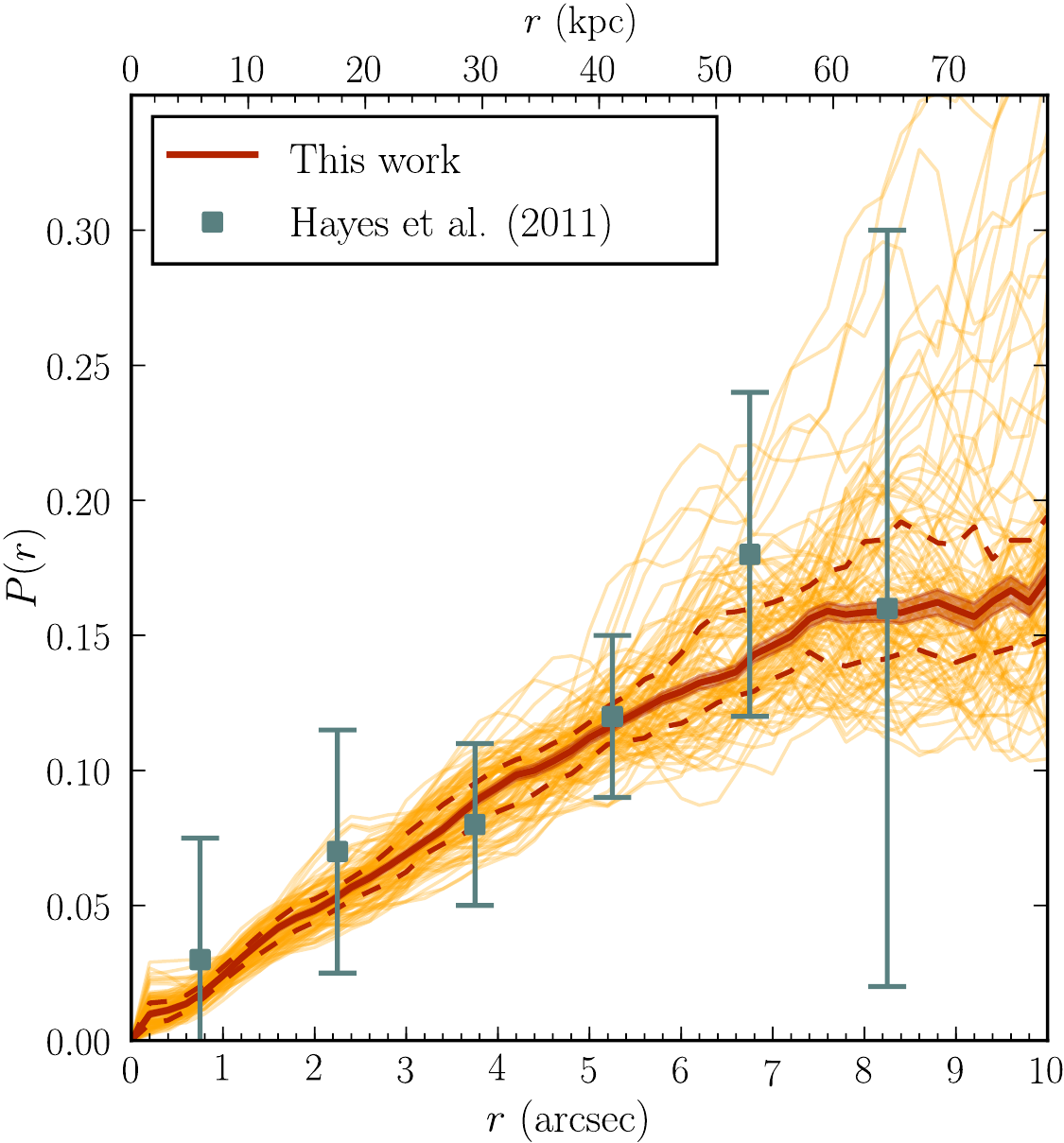}
      \put(80,15){\large\textbf{(c)}\protect\label{fig:profiledir_both}}
    \end{overpic}
  }%

  \caption{Polarization radial profiles for (a) extragalactic emission, (b) galactic emission, and (c) the overall \lya emission. The thin, orange lines show the profile corresponding to each LOS; the solid, red line is the median profile; the dispersion along different LOS is represented by the two dashed, red lines (first and third quartiles). The red area show the $3\sigma$ confidence limits inferred from our bootstrap experiment (see Appendix \ref{sec:bootstrap}). The data points are taken from \Hayes.}
  \label{fig:profiledir}
\end{figure*}

In order to compare our simulation to \Hayes polarimetric observations, we need to characterise both the direction and the degree of polarization. To describe the latter, we compute radial profiles for the different components of the \lya emission. Figure \ref{fig:profiledir} displays the polarization profiles obtained for each component of the signal: emission from extragalactic gas (panel \ref{fig:profiledir_gas}), \lya photons produced in the star-forming ISM (panel \ref{fig:profiledir_stars}), and the combination of the two (panel \ref{fig:profiledir_both}). We compare these results with \Hayes observations, displayed as filled squares with errorbars.

The main result of our study is that the polarization profile produced only by the extragalactic gas rise up to 15\%, similar to what is observed by \Hayes. This is unexpected: in this non-idealised setup, the extended emission does not wash out the polarization.
This is mainly because the gas distribution is not homogeneous. Even if we refer to the extragalactic emission as an extended source, it is still much more concentrated in the inner region of the blob, as can be seen in the SB profiles of Fig.~\ref{fig:sbgasstars}.
\citet{Dijkstra2009} suggested that the low volume filling factor of the cold streams would prevent the polarization to arise: there is indeed only little chance that a photon that has escaped from a filament will encounter another one before being observed. However, we found that the photons responsible for the polarization signal mostly travel radially outwards inside the gas, and then escape their filament at the last scattering (see Sec.~\ref{sec:origin}).

Furthermore, if we look at the galactic component only, it is clearly inconsistent with \Hayes observations: the polarization profile is too steep in the central region, meaning that it is compulsory to take the extended emission into account. We checked that this is not an artifact resulting of the choice of the pixelization. While the profiles presented in Fig.~\ref{fig:profiledir} corresponds to maps with a pixelization ($\gtrsim 0.12\arcsec{}$) much finer than the spatial resolution of the observations, we verified that our results hold for a coarser spatial resolution ($\sim 1.25 \arcsec$). In the experiment with larger pixels, we noted a decrease of the degree of polarization at large distance ($\gtrsim 5\arcsec$), but not strong enough to alter our results.

\subsubsection{Polarization angle}
\label{sec:polangle}

\begin{figure}
  \centering
  \resizebox{.8\hsize}{!}{\includegraphics{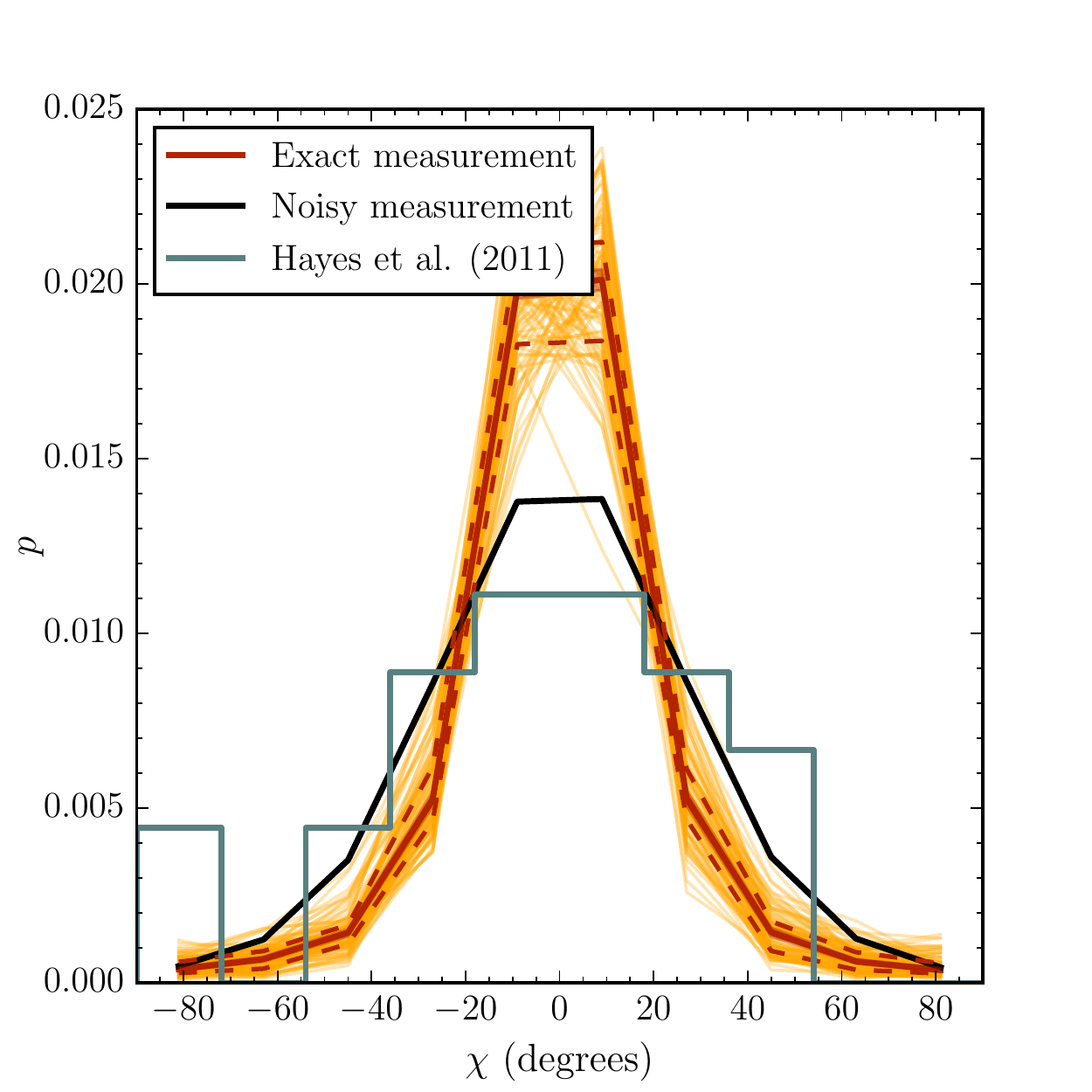}}
  \caption{Distribution of the angle between polarization and tangential direction. The thin, orange lines show the profile corresonding to each LOS; the solid, red line is the median profile; and the two dashed, red lines denotes the first and third quartiles. The red area show the $3\sigma$ confidence limits inferred from our bootstrap experiment (see Appendix \ref{sec:bootstrap}), and the teal line is the distribution taken from \Hayes. In black, we show the profile we obtain by assuming some noise in the angle measurement.}
  \label{fig:anglehist}
\end{figure}

The second observed attribute we can produce is the polarization angle. Qualitatively, the direction of polarization in a given pixel of the map seems to be aligned in circles around the centre of the blob, as shown on Fig.~\ref{fig:illustration_results}. A more quantitative study can confirm this: for each pixel on the map, we compute the difference between the polarization angle and the tangential angle. We then rebin the resulting distribution to match \Hayes bins, and the result is shown on Fig.~\ref{fig:anglehist}. The polarization angle is not random at all, but rather aligned with the tangential angle.

This is qualitatively compatible with the results of \Hayes: they also find a clustering of the values around zero. A more quantitative comparison shows that our distribution is much more peaked around zero. However, in our numerical experiment, we have no measurement error on the polarization angle in each pixel, which is not true in the case of observations. We assumed a gaussian error of width $20 \degr$ on the angle measurement, and recomputed the angle distribution. The result, shown as the black curve on Fig.~\ref{fig:anglehist}, is in much better agreement with the observations.

\subsubsection{Origin of the polarization}
\label{sec:origin}

\begin{figure*}
  \centering
  \subfloat{%
    \begin{overpic}[width=7.5cm]{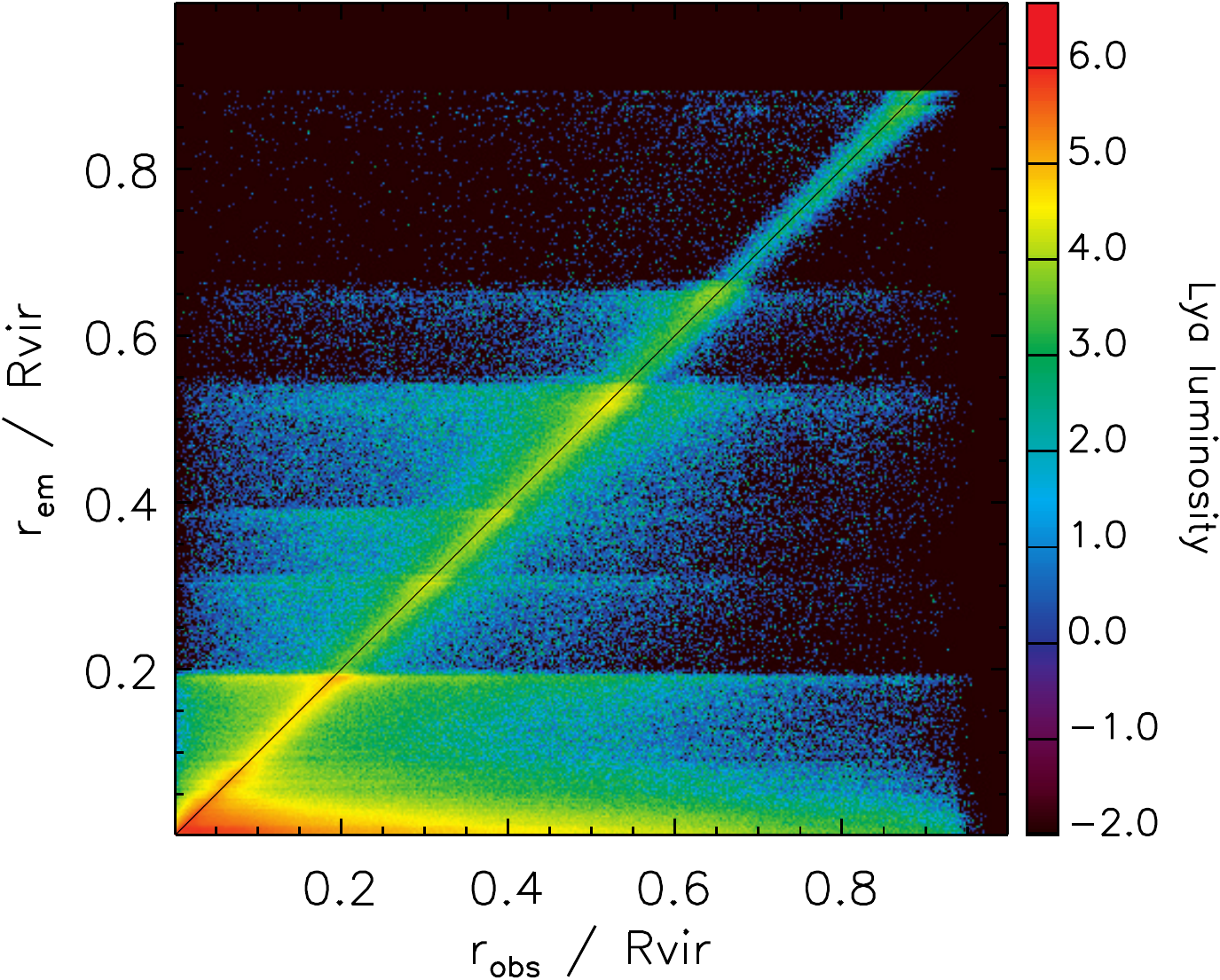}
      \put(5, 75){\Large\textbf{(a)}\protect\label{fig:rproj_remproj_stars}}
    \end{overpic}
  }%
  \hspace{2em}
  \subfloat{%
    \begin{overpic}[width=7.5cm]{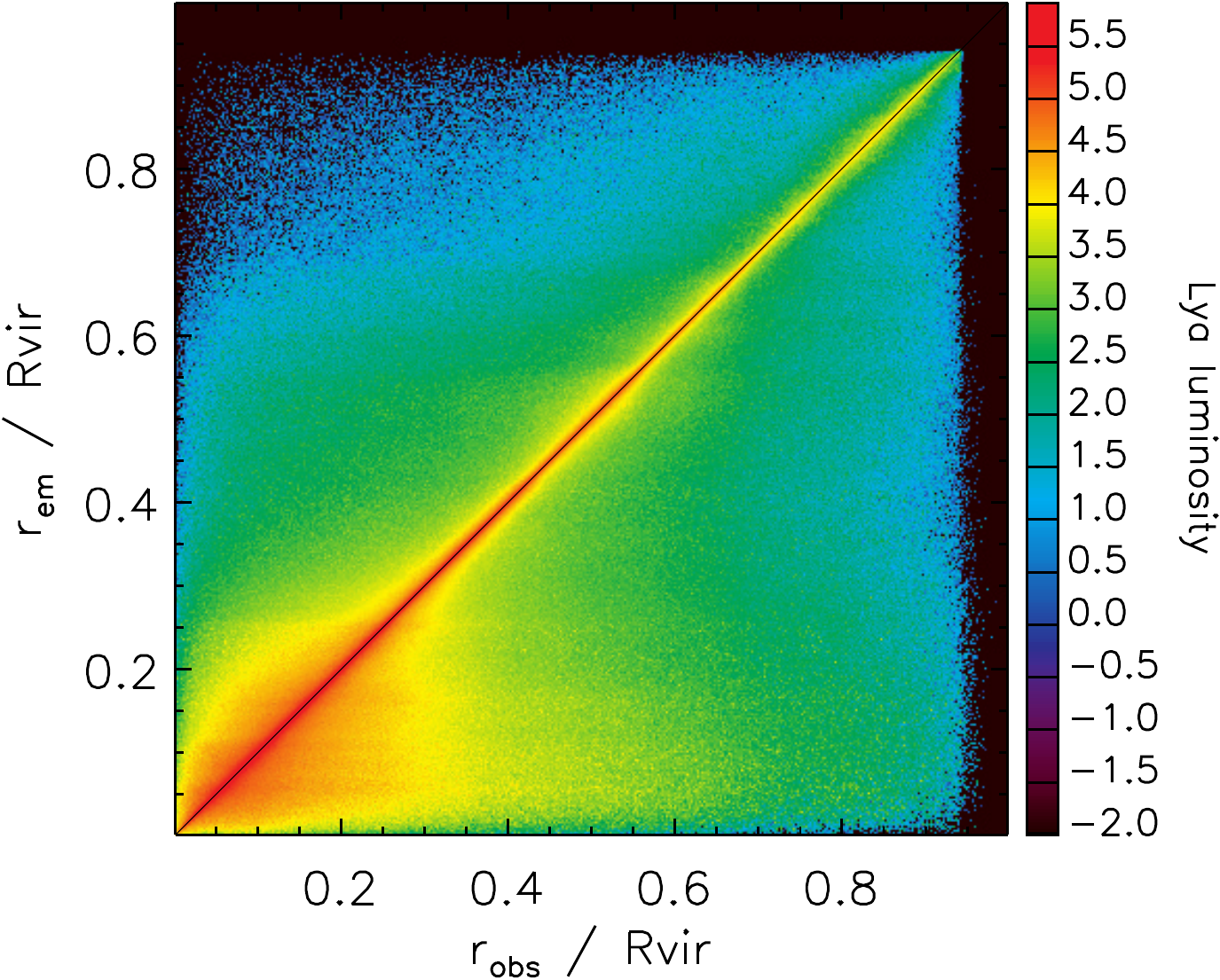}
      \put(5, 75){\Large\textbf{(b)}\protect\label{fig:rproj_remproj_gas}}
    \end{overpic}
  }%
\caption{Distribution of the radii of emission, $r_{\rm em}$, for each observation radius, $r_{\rm obs}$, weighted by the (normalised) luminosity of the photons. These radii are not distances to the halo centre, but projected on the observation plane perpendicular to the line of sight for each simulated photon. (a) the galactic component, (b) the extragalactic component. }
\label{fig:robsrem}

\end{figure*}

Polarization is a geometrical effect, which arises naturally in a configuration with centrally concentrated emission which is scattered outwards. From our numerical experiment, we find that extended extragalactic \lya emission generates a polarized nebula with a relatively strong polarization signal (15\% close to the virial radius). This polarization emerges for the same reason: photons statistically scatter outwards before being observed. 

On Fig.~\ref{fig:robsrem}, we show the 2D histogram (weighted by luminosity) of the projected emission radius $r_{\rm em}$ (where the MC photons are cast) as a function of the projected observed radius $r_{\rm obs}$ (where the MC photons last scatter before being observed). These projected radii are projections onto the plane perpendicular to the direction of propagation of each MC photon. The left (resp. right) panel shows the distribution of projected $r_{\rm em}$ versus projected $r_{\rm obs}$ for the galactic (resp. extragalactic) emission. The prominent feature in both cases is the diagonal line, showing that a significant part of observed photons escape close to their emission site even in the case of galactic emission.  The asymmetry between the upper and lower half planes illustrates that more photons escaping at a given $r_{\rm obs}$ where emitted at a smaller radius. There is a strong (expected) asymmetry for the galactic emission, and a lighter but noticeable asymmetry in the extragalactic case. This explains the steeper polarization profile for the galactic sources than for the extragalactic sources (see Fig.~\ref{fig:profiledir}). Each horizontal feature on the left panel corresponds to the location of a \lya\ source (satellite galaxy), and illustrates the fact that a significant fraction of photons emitted by external sources (with large $r_{\rm em}$) scatter also on the central parts of the blob, and escape at smaller $r_{\rm obs}$.  To summarize, for a significant fraction of the \lya MC photons (more than 55\% of the extragalactic luminosity), the emission location is close to the last scattering place. Those \lya photons do not contribute to the observed polarization. 

\begin{figure}
  \centering
  \resizebox{.8\hsize}{!}{\includegraphics{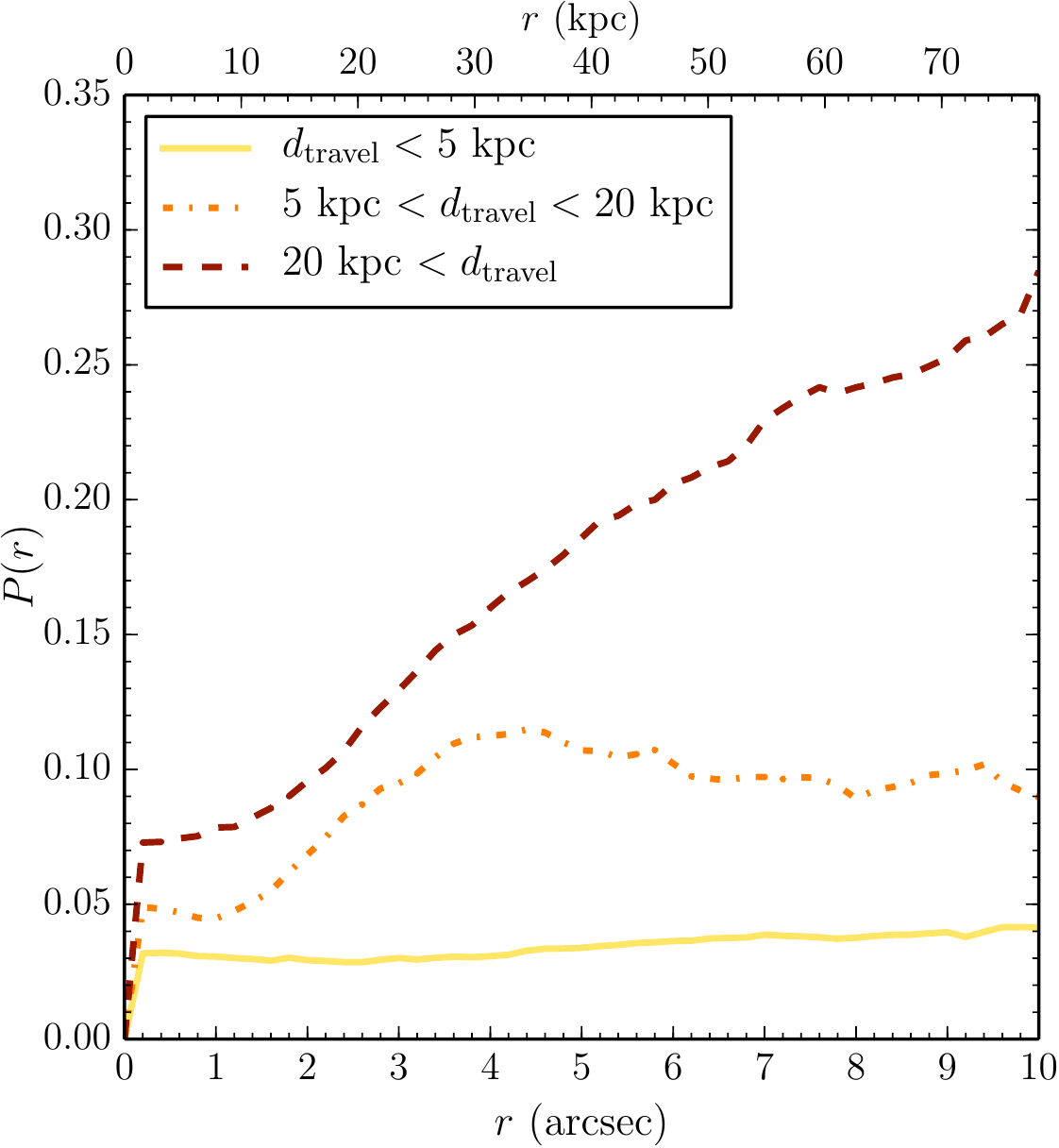}}
  \caption{Polarization profile obtained by selecting photons according to the distance between the emission and the last scattering.}
  \label{fig:dop_origin}
\end{figure}

To sketch this out in a more quantitative manner, we now focus on the extragalactic component (i.e. cooling radiation from the gas). In Fig.~\ref{fig:dop_origin}, we show the polarization signal due to extragalactic MC photons which have traveled less than 5 kpc (resp. between 5 and 20 kpc, and more than 20 kpc) as the yellow (resp. dot-dashed orange, and dashed red) curve. The photons that travel more are the ones responsible for the polarization. Note that by selection, they do tend to come from the central regions as well.

\subsection{Scattering in the IGM}
\label{sec:IGM}

So far, we limited our analysis to the photons scattered within the virial radius of the halo, thus assuming that the effect of the IGM was negligible. Previous works, e.g. by \DL showed however that for a galaxy without strong outflows (as it is the case in our simulation), radiation scattered in the IGM will carry a low polarization level, typically around 2\%, and has a very flat surface brightness profile. This is because as they travel through the IGM, \lya photons will be blueshifted and could experience a significant number of scattering, which would reduce the level of polarization.

\begin{figure}
  \centering
  \resizebox{\hsize}{!}{\includegraphics{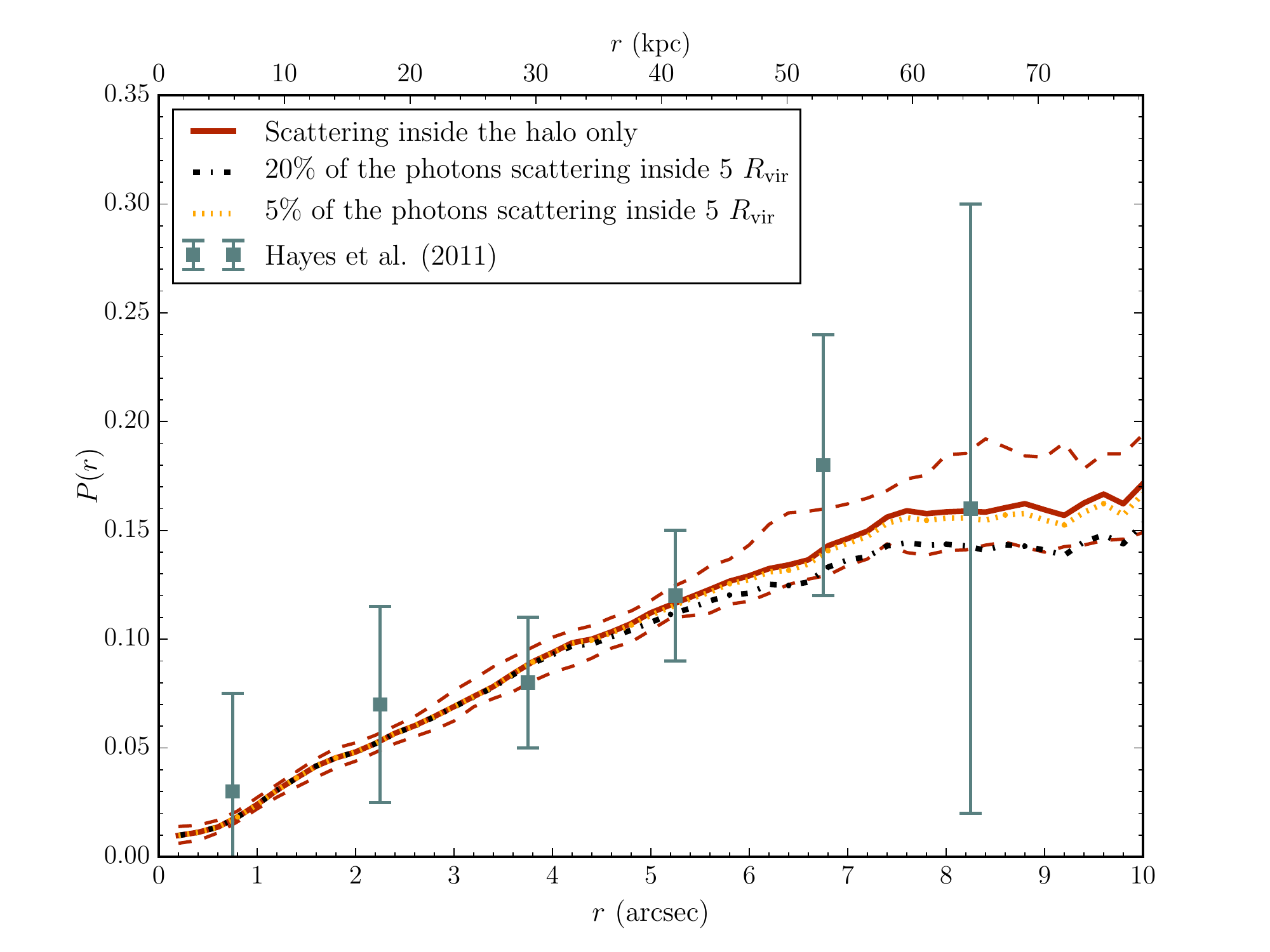}}
  \caption{Effect of the IGM on the polarization profile. Compared to the profile right outside of the halo (solid red line), scattering in the IGM (dash-dotted and dotted black lines) reduces slightly the degree of polarization at large radii, but not much (if we redistribute the photons in a large enough area).}
  \label{fig:IGM}
\end{figure}

While we cannot fully describe the \lya{} resonant scattering in the IGM with the current version of \mclya (a volume larger than currently investigated would not fit in the computer memory), we still can get an idea of to what extent taking further scattering into account would affect our results. We denote by $f$ the fraction of the luminosity that will scatter in the IGM. To compute the value of $f$, we assume that photons escaping the halo redwards of the lya{} line will be observed directly, and that a fraction of $1 - \mathcal{T}_{\rm IGM}$ of the blue photons will undergo further scattering, with $\mathcal{T}_{\rm IGM} \simeq 0.67$ being the mean IGM transmission at $z \sim 3$ \citep[see e.g.][]{Faucher-Giguere2008, Inoue2014}. We can compute the value of $f$ from the spectrum averaged over all directions : a fraction $f_b \sim 60.5\%$ of the photons are bluewards of $\lambda_\lya$, resulting in $f = f_b \times (1-\mathcal{T}_{\rm IGM}) \simeq 0.2$, meaning that on average, approximately 20\% of the photons in our simulation will be scattered in the IGM. While this is only a first order approximation, it gives a reasonable estimate of the amount of photons that will be scattered in the IGM. We discuss its validity in Appendix~\ref{sec:app:IGM}.
Following the findings of \DL that scattering in the IGM results in a rather flat profile, and to maximise the effect, we uniformly redistribute the total luminosity contributed by these photons in a patch of sky of 10 \Rvir on a side, such that the photons have travelled $\sim$ up to 5 \Rvir, corresponding to an area larger than the maps of Fig.~\ref{fig:illustration} by a factor of 25.
We assign to these photons a degree of polarization of 2\%, following the results of \DL, and assume that the linear polarization follows the same pattern around the galaxy as before. More precisely we compute the $I_{\rm IGM}$, $Q_{\rm IGM}$ and $U_{\rm IGM}$ maps as
\begin{align}
  \label{eq:IQUigm}
  \begin{split}
    I_{\rm IGM} &{}= f \times I_{\rm direct} / A ,\\
    Q_{\rm IGM} &{}= P_{\rm IGM} \times \frac{I_{\rm direct}}{\sqrt{1 + \alpha^2}} {\rm sgn}(Q_{\rm direct}),\\
    U_{\rm IGM} &{}= \alpha \times Q_{\rm IGM},
  \end{split}
\end{align}
where $A = 25$ is the dilution factor due to the larger area over which photons are redistributed, $P_{\rm IGM} = 0.02$ is the polarization level of the radiation scattered in the IGM, and $\alpha = \frac{U}{Q}$. We then sum the direct and scattered contributions as
\begin{align}
  \label{eq:resumIQU}
  \begin{split}
    I &{}= I_{\rm IGM} + (1-f) I_{\rm direct},\\
    Q &{}= Q_{\rm IGM} + (1-f) Q_{\rm direct},\\
    U &{}= U_{\rm IGM} + (1-f) U_{\rm direct}.
  \end{split}
\end{align}
We present the results of this experiment on Fig.~\ref{fig:IGM}. The red lines are the same as in Fig.~\ref{fig:profiledir}. The dash-dotted black line shows the polarisation profile we obtain with the above calculation, assuming that all the photons scattered by the IGM beyond \Rvir are seen as coming from within a extended surface of side $10\ \Rvir$. We find that quantitatively, the effect is small, and that the signal remains within the error bars of Fig.~\ref{fig:profiledir}. More importantly, the deviation occurs are large radii, and the IGM has no effect on the signal within $\sim 40$ kpc where the constraints are stronger.
We note that this is likely to overestimate the impact on the IGM on the polarisation profile. Based on the work of \citet{Laursen2011}, we estimated that about 5\% of the photons crossing \Rvir will be scattered within 5 \Rvir (compared to 20\% scattered in total, see Appendix~\ref{sec:app:IGM}). This means that most photons will scatter very far away from the source.
This implies that the luminosity contributed by those photons will be diluted over a much larger area. 
On Fig.~\ref{fig:IGM}, the orange dotted curve shows the more realistic polarisation profile that we obtain when we only redistribute these 5\% of the luminosity within an area of 10 \Rvir on a side. It is barely distinguishable from the model withough IGM (see Appendix~\ref{sec:app:IGM}).
Keeping in mind that our model is only a first order approximation, it seems that scattering of \lya{} radiation well outside the virial radius is not likely to alter dramatically our polarization profiles. Strictly speaking, though, all the previous results on the polarization should be regarded as upper limits.

%__________________________________________________________________
\section{Discussion and conclusions}
\label{sec:conclusions}

In this work, we have performed \lya radiative transfer through a LAB simulation, previously discussed by \RB. We considered separately the galactic and extragalactic contributions to the \lya luminosity, and we followed the polarization of \lya photons during their journey through the blob.

% SUMMARY
Our main results are the following:
\begin{itemize}
  \item[\textbullet] We confirm that the results of \DL for their idealised ``cooling model'' holds in the case of a more complex but realistic distribution of gas: the cooling radiation produces a polarized signal.
  Furthermore the polarization radial profiles computed by only taking the extragalactic contribution into account is compatible with observational data.
  \item[\textbullet] A \lya escape fraction of the galactic contribution of 5\% is enough to find a good agreement with \Hayes results. This means that a non-negligible extragalactic contribution to the luminosity is compatible with current polarimetric observations.
  \item[\textbullet] The (galactic) contribution of the star formation to the \lya luminosity of a HzLAN with no associated AGN is small, and the impact of the scattering on the SB profile is similar to the effect of a convolution with a PSF. This confirms that the extent of the HzLANs presented in \RB is correct.
\end{itemize}

% LIMITATIONS
It is important to stress some of the potential limits of our investigation. First, the galactic contribution is uncertain in our simulation because of the under-resolved structure of the ISM. Nevertheless, our estimation of the star formation rate ($SFR \simeq 160 \Msun.\mbox{yr}^{-1}$, shared between all the galaxies in the halo) is typical for giant LABs \citep{Fardal2001}. The total galactic contribution to the \lya luminosity is however the product of the intrinsic galactic luminosity by the \lya escape fraction $f_{\mathrm{esc}}$. Our results are consistent with \Hayes data with a typical value of $f_{\mathrm{esc}} = 5\%$ \citep{Garel2012}.
In our model, the spectral shape of the stellar component of the \lya emission can be arbitrarily selected. However, we have tested that the input spectrum of the stellar component has little impact on the SB and polarization profiles, provided the choice of the spectrum is physical enough (gaussian profile, P-Cygni like profile). Further work would be needed to produce strong predictions for spectro-polarimetric studies. We also assume that the \lya photons are isotropically emitted from the galaxies. From \citet{Verhamme2012}, we know that this is not true. However, since (i) we have several galaxies in the halo and (ii) the photons scatter a lot in the CGM, there should be no favoured escape direction. This is corroborated by the low polarization degree in the inner regions of the blob.
Another possible issue is that the simulation we use is somewhat idealised: \RB halo includes neither cooling below $10^4\ \mbox{K}$, nor supernova feedback. Adding these ingredients could potentially alter the structure of the central region blob (CGM and inner parts of the streams), and that is precisely where most of the extragalactic gas contribution \lya emission comes from. Further work will be needed to carefully quantify the impact of cooling and feedback on our results.
Finally, we must note that we only take into account scattering within the virial radius, leading to a most likely small overestimation of the degree of polarization we compute, at large radius.
Lastly, we must be aware of the lack of statistics on HzLANs polarimetric studies. There are currently only two positives observations of polarization around a giant HzLANs \citep{Hayes2011, Humphrey2013}, and we have no certainty that our blob is a perfectly typical giant blob. However, the stability of the polarization and SB profiles when changing the LOS is reassuring. Once again, further work is needed to address this question.

% Connection to radio loud HzLAN
In this work, we have only considered two possible sources for the \lya radiation: cooling radiation emitted by the accretion-heated gas, and \lya emission from the \hii star-forming regions. We have not investigated the possibility that the gas is ionised by central AGN. \citet{Overzier2013} suggested that virtually all the most luminous HzLANs are associated with AGNs, and that the for the less luminous HzLANs, the central black-hole is just not in an ``active'' state. This is compatible with the scenario of \citet{Reuland2003}. In this picture, HzLANs are the signatures of the first stage of the building of massive galaxies. As the gas falls onto the halo, it dissipates its energy via \lya cooling, producing a blob. As the gas accretes, stars and galaxies begin to form and merge, triggering at some point the central AGN.

The \lya polarization radial profile arising from a non idealised distribution of gas appears finally as a rich and complex tool. This work is a first step towards a better understanding of \lya polarimetric observations. However, another step needs to be done to use them to study the relative contributions of extragalactic versus star-formation channels of \lya production.

%__________________________________________________________________

\begin{acknowledgements}
The authors would like to thank Matthew Hayes and Claudia Scarlata for stimulating discussions, and the anonymous referee for insightful comments. We wish to thank Gérard Massacrier and Stéphanie Courty for valuable help at the early stages of the project. We are also grateful to Léo Michel-Dansac whose help with the computing centres was priceless. 

JR acknowledges financial support from the Marie Curie Initial Training Network ELIXIR of the European Commission under contract PITN-GA-2008-214227, the European Research Council under the European Union's Seventh Framework Programme (FP7/2007-2013) / ERC Grant agreement 278594-GasAroundGalaxies, and the Marie Curie Training Network CosmoComp (PITN-GA-2009-238356).

The initial simulations of the blob were performed using the HPC resources of CINES under the allocation c2012046642 made by GENCI (Grand Equipement National de Calcul Intensif), and all the other ones were performed using the computing resources at the CC-IN2P3 Computing Centre (Lyon/Villeurbanne - France), a partnership between CNRS/IN2P3 and CEA/DSM/Irfu.

\end{acknowledgements}

\bibliographystyle{aa}
\bibliography{blobpolarization}

\begin{appendix}
\section{Test of the code}
\label{sec:test}
To test the validity of our polarized \lya transfer code, we compare it to the first case considered by \DL, \lya scattering on galactic superwinds. This is pictured with a simple toy model: a thin spherical shell of column density $N_\hi = 10^{19}\,\mbox{cm}^{-2}$ or $10^{20}\,\mbox{cm}^{-2}$  illuminated by a central source. The radius of the shell is $10\,\mbox{kpc}$, and the expansion velocity is $200\ \mbox{km}.\mbox{s}^{-1}$. We use a Gaussian profile with a width corresponding to a temperature of $T = 10^4\ \mbox{K}$ for the input spectrum of the \lya emission.

In their code, \DL model gas by concentric shells of given densities and velocities, whereas our code use an AMR grid to describe the gas. To create the shell, we fill an unrefined grid with $512^3$ cells using Monte Carlo integration with $10^8$ points.

\begin{figure}[h]
  \centering
  \resizebox{.8\hsize}{!}{\includegraphics{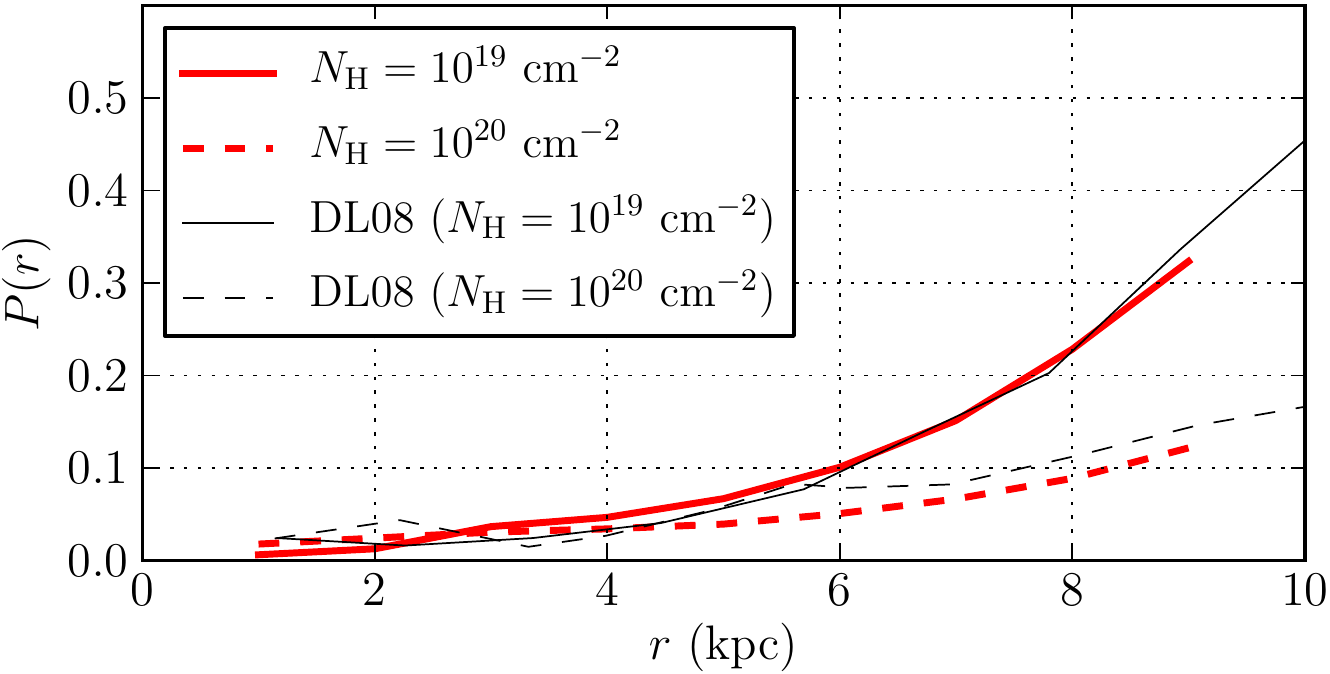}}
  \caption{Polarization profile for an expanding spherical shell. Results from \DL are shown in black, and our results are shown in red. The solid (dashed) line corresponds to a column density of $N_{\mathrm{H}} = 10^{19}\ \mbox{cm}^{-2}$ ($N_{\mathrm{H}} = 10^{20}\ \mbox{cm}^{-2}$).}
  \label{fig:shelltest}
\end{figure}

We then cast and follow $10^6$ Monte Carlo photons in this setup, in order to produce polarization profile using the method described in Sec.~\ref{sec:observation}. On Fig.~\ref{fig:shelltest}, we compare our results (in red) to the profile of \DL (black). We find a fairly good agreement, both for the $N_{\mathrm{H}} = 10^{19}\ \mbox{cm}^{-2}$ shell (solid line) and for the $N_{\mathrm{H}} = 10^{20}\ \mbox{cm}^{-2}$ shell (dashed line).

\section{Impact of the sampling}% of the \lya emission}
\label{sec:bootstrap}

\begin{figure*}
  \centering
  \subfloat{%
    \begin{overpic}[width=6.5cm]{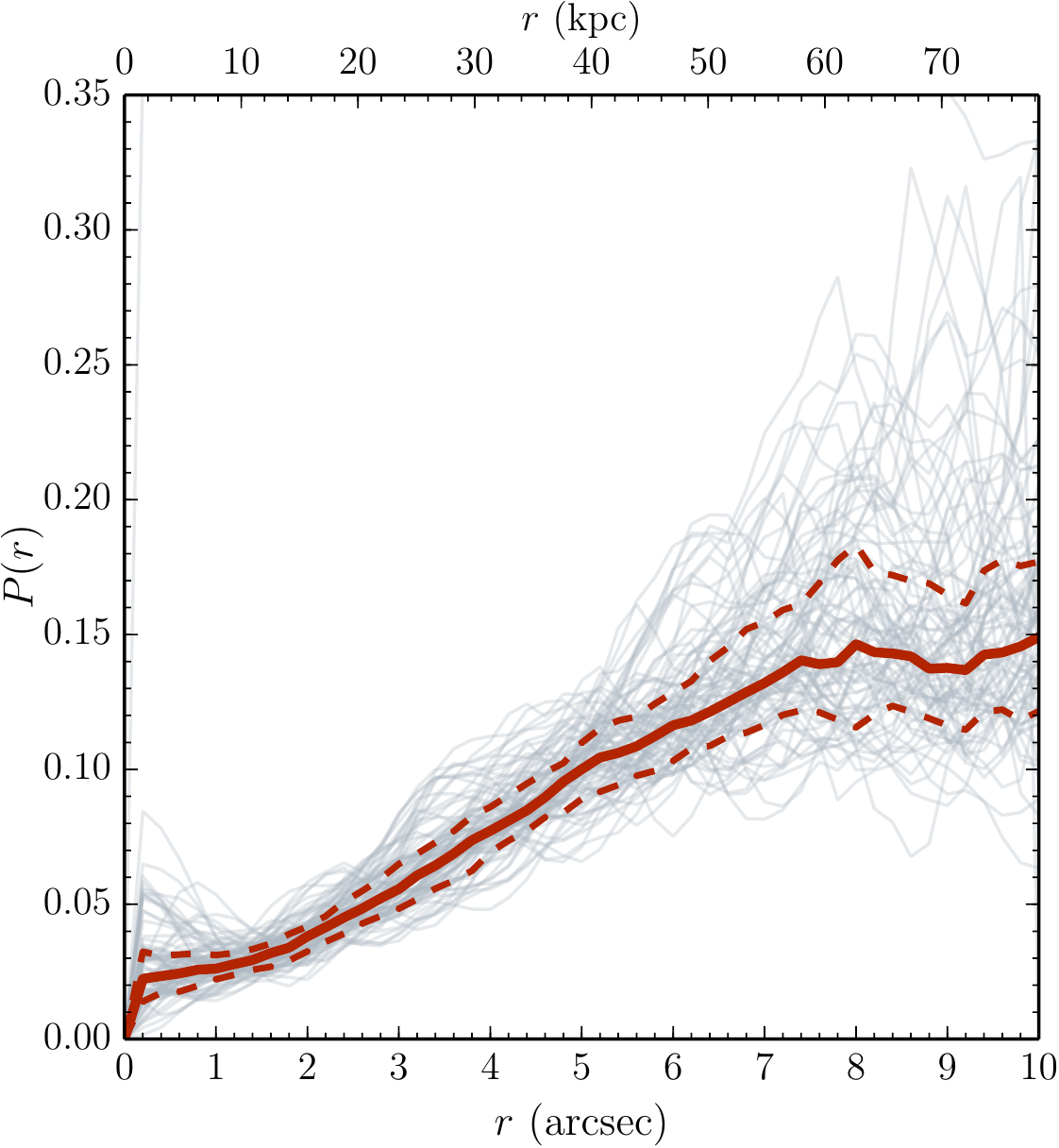}
      \put(80, 15){\large\textbf{(a)}\protect\label{fig:bootstrap_gas_onesample}}
    \end{overpic}
  }%
  \subfloat{%
    \begin{overpic}[width=6.5cm]{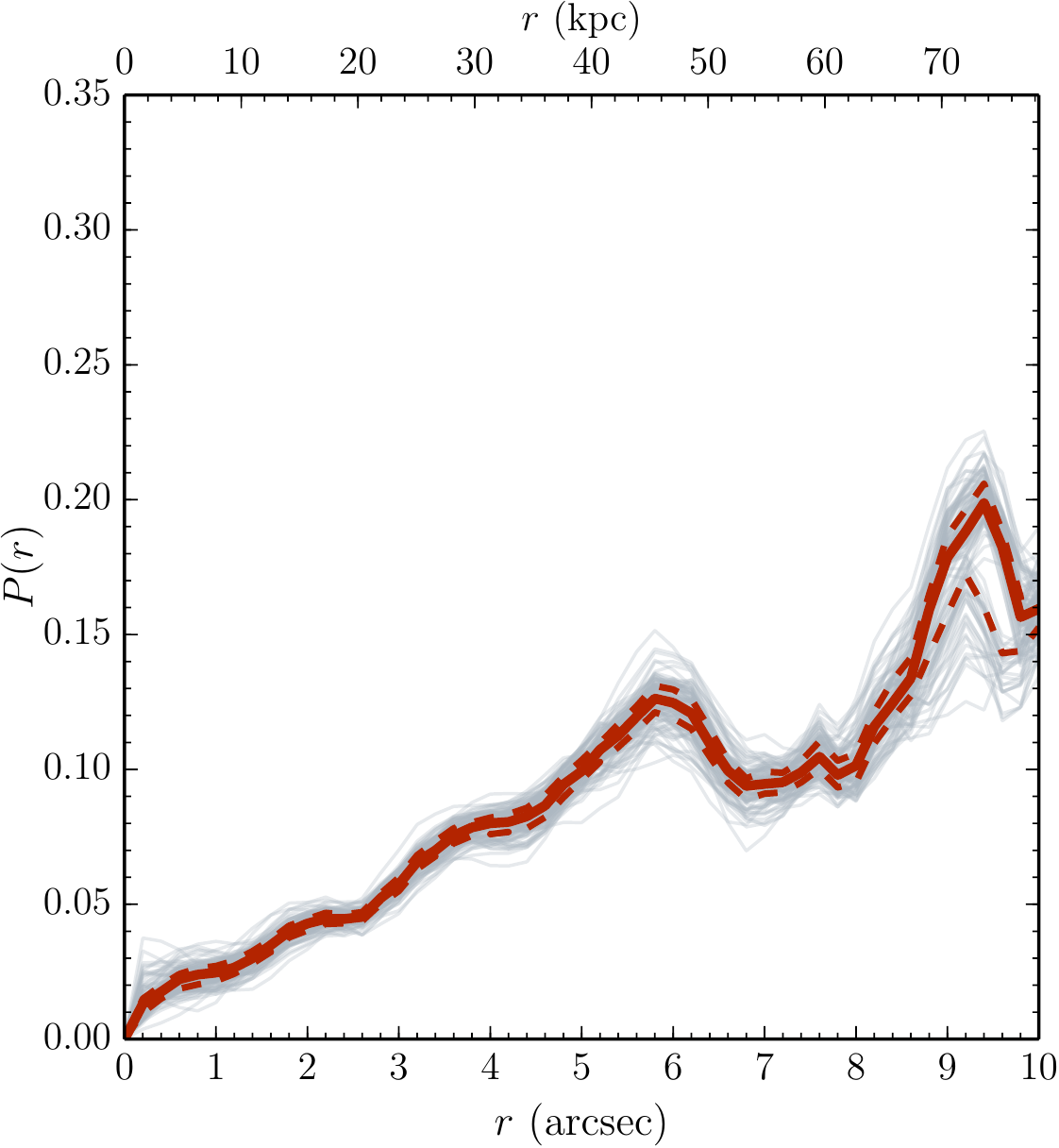}
      \put(80, 15){\large\textbf{(b)}\protect\label{fig:bootstrap_gas_onedir}}
    \end{overpic}
  }%
  \\
  \subfloat{%
    \begin{overpic}[width=6.5cm]{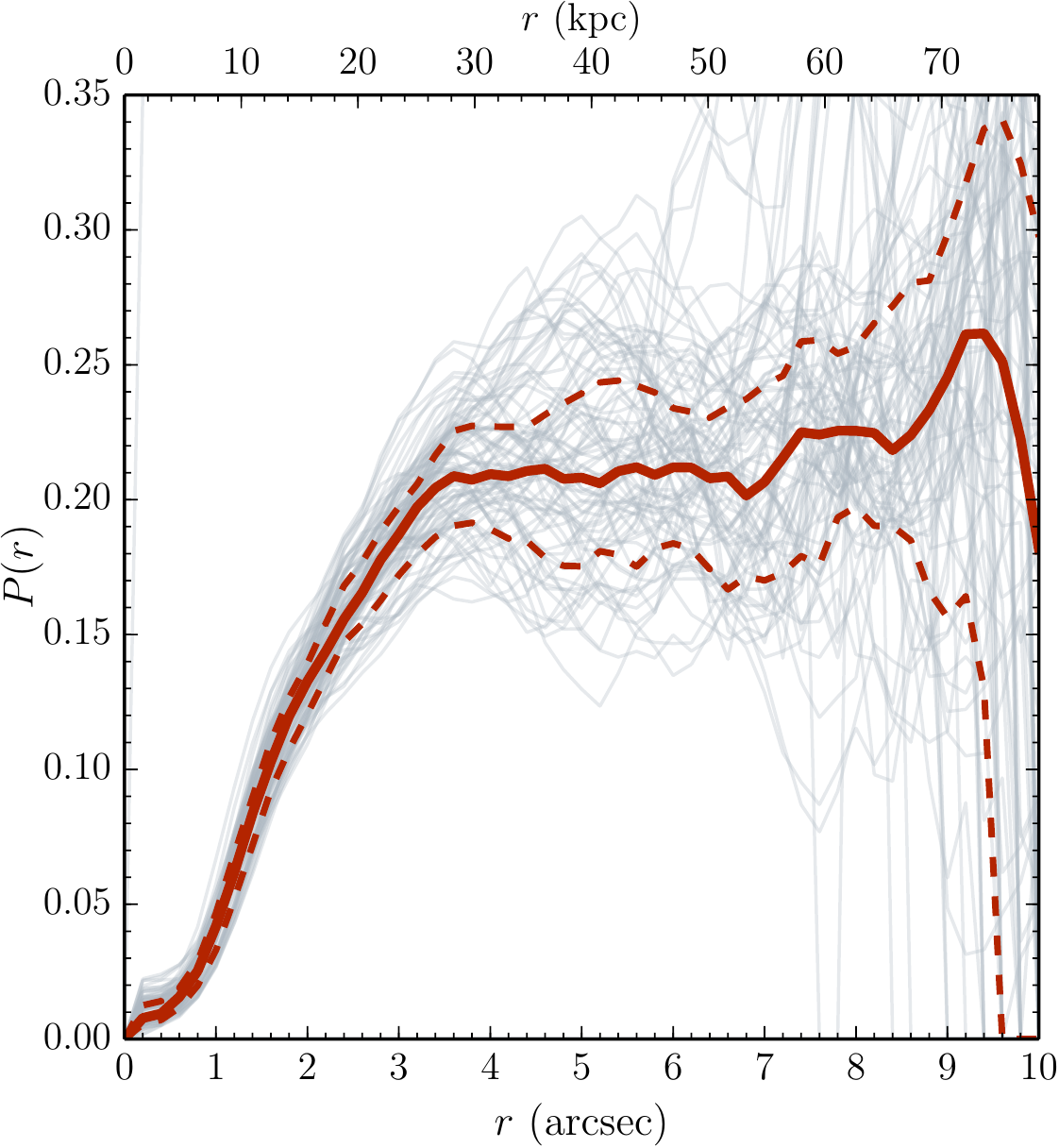}
      \put(80, 15){\large\textbf{(c)}\protect\label{fig:bootstrap_stars_onesample}}
    \end{overpic}
  }%
  \subfloat{%
    \begin{overpic}[width=6.5cm]{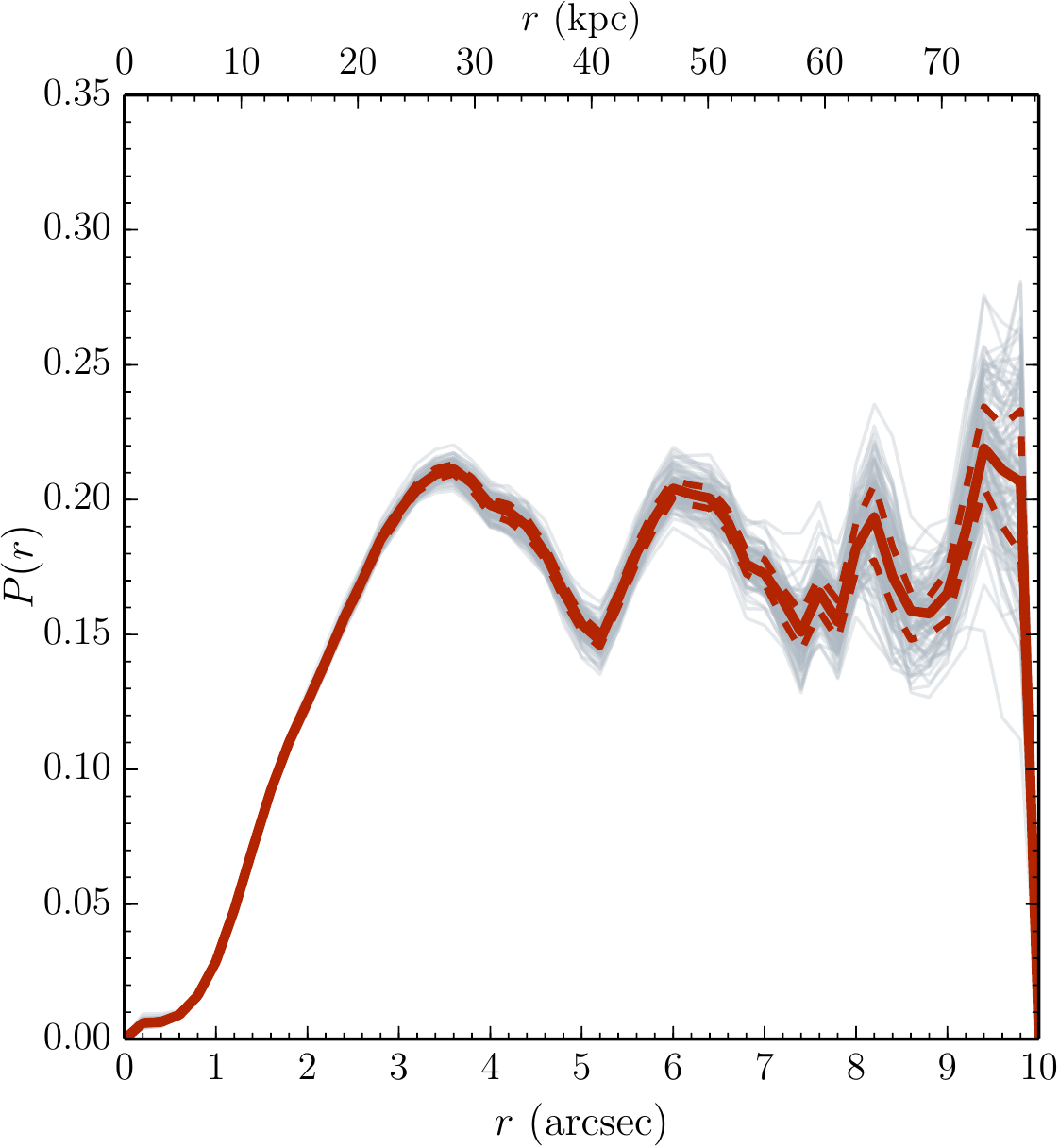}
      \put(80, 15){\large\textbf{(d)}\protect\label{fig:bootstrap_stars_onedir}}
    \end{overpic}
  }%

  \caption{Polarization profile for different subsets of photons. In all the plots, the red, solid line is the median profile; the red, dashed lines are the first and third quartiles. The thin, gray lines represent the profiles for (a) 100 LOS with the same photons subset for the gas emission; (b) 100 different photons subsets seen with the same LOS for the gas emission; (c) 100 LOS with the same photons subset for the galactic emission; (b) 100 different photons subsets seen with the same LOS for the galactic emission.}
  \label{fig:bootstrap}
  
\end{figure*}

In Sec.~\ref{sec:gaslya}, we mentioned that we sampled the \lya emission from the gas nebula with 150 Monte Carlo photons per simulation cell. With only 150 photons, it is impossible to get a proper description of all the physics of the \lya emission in each cell. Indeed, we need to sample not only the initial position of the photon within the cell (three degrees of freedom), but also the initial propagation direction (three degrees of freedom), the initial frequency (one degree of freedom) and the initial polarization direction (two degrees of freedom). It is impossible to sample this 9D parameter space with only 150 points. The same argument holds for the galactic emission: the initial position is well sampled (the number of photons is largely greater than the number of sources), but we still have to sample the six other variables. 

To ensure that our results results are not just a statistical artefact, we need to assess the robustness of our method and to understand how the (under)sampling might affect our results.
For this purpose, we used a bootstrap method, both for the emission from the gas and from the star particles. In the case of the gas emission, for each simulation cell, we performed the analysis presented in Sec.~\ref{sec:results} with a subset of 120 photons (80\%) randomly selected. Instead of selecting the random subset \emph{per cell}, we could have selected \emph{80\% of all} the photons. However, this experiment would not answer the question of the effect of undersampling \emph{each cell}.
We did this for 100 different randomly selected subsets. For the star particles, we selected random subsets with 80\% of all the photons, since the number of photons emitted per individual star particle is very large, and sampling the initial position of the photon is not a problem. From this, we get 100 different profiles (either for polarization or for surface brightness), and the standard deviation $\sigma_{\mathrm{boot}}$ of this set of profiles gives an estimate of the error caused by the undersampling of the parameter space. This estimate (as $3\sigma_{\mathrm{boot}}$) is displayed on our profiles in the main text of the paper as red, semi-transparent areas.

The results are displayed on Fig.~\ref{fig:bootstrap}. The top (bottom) panels show the results of our bootstrap experiment for the extragalactic (galactic) emission. The panels \ref{fig:bootstrap_gas_onesample} and \ref{fig:bootstrap_stars_onesample} show 100 polarization profiles corresponding to 100 LOS for one of the subsets as thin, gray lines (for the extragalactic and galactic emission, respectively). The panels \ref{fig:bootstrap_gas_onedir} and \ref{fig:bootstrap_stars_onedir} show 100 profiles corresponding to 100 subsets for a given LOS. Following the convention used throughout this paper, the solid, red line is the median profile and the two dotted lines shows the interquartile range.

It is reassuring to note that the variation over the LOS is much more important than the variation between photons subsets. This means that our sampling of the \lya emission has much less impact on the observed polarization profile that the choice of the LOS.

\section{Selection effects}
\label{sec:angle}

Because of our limited number of photons, we need to select photons in a (small) cone around each line of sight. This is an approximation, and might change the polarization properties we get from our analysis. To test this, we performed the same analysis as before but changing the angular opening of the cone, as well as the minimum number of photons selected to compute the polarization properties in a pixel.

Although we would ideally prefer to estimate the polarization in small beams ($\sim$ few arcsec), our data does not allow us to measure the polarization signal in beams of less than $3 \degr$. Below these scales, the (strong) polarization degree is dominated by noise, and its large-scale coherence disappears. We verified this by performing the analysis with an opening angle of $15 \degr$, and selecting randomly a small fraction of these photons. The fraction corresponds to $\frac{\Omega (1\degr)}{\Omega (15 \degr)} \simeq 4\%$. The resulting profile is very similar to the one we obtain by reducing the opening angle to $1\degr$, meaning that the dominant effect here is not the error due to the selection in a cone, but rather the limited number of photons. However, at larger opening angles (from $\sim 3$ to $60 \degr$), we consistently find the same profile as shown in Fig.~\ref{fig:profiledir}, with small deviations of less than 10\%. This suggests that our results, which are robust at $15 \degr$, can also be compared to observations made with much smaller beams. 

% One last test that we performed was to run again the analysis, but this time using the photons proper direction when projecting and computing polarization properties, rather than the line of sight direction. This setup is not exactly the same as before: indeed, the polarization plane will not be the same for each photon. This resulted in a slightly lower polarization profile, of typically 5\% of the absolute value.

The surface brightness profile is much more robust, and is mostly unaffected by these experiments. Even with an $1\degr$ cone, the relative error is smaller than 10\%.

\section{Scattering in the IGM}
\label{sec:app:IGM}

In Sect.~\ref{sec:IGM}, we assumed that $1 - \mathcal{T}_{\rm IGM} = 1 - 67\% \sim 33\%$ of the photons bluewards of $\lambda_\lya$ would be subject to further scattering during their journey through the IGM. In this section, we try to estimate more carefully the impact of the IGM on the \lya photons escaping the halo.

We use the results of \citet{Laursen2011}, who followed the transfer of radiation in the IGM in the vicinity of the \lya line using a large cosmological simulation. In their Fig.~11, they present the transmission function of the IGM at various redshifts and for sightlines originating at various distances from the centre of their simulated galaxies.
As we fully perform the \lya RT only up to the virial radius of our blob, we need to use their transmission function as a proxy for the actual transfer through the IGM.
We extracted the curves corresponding to sightlines originating at the virial radius for both $z=2.5$ and $z=3.5$. We fitted the data using the following ad-hoc function $T(\lambda) = T(\lambda; \lambda_0, T_b, T_{\rm min}, \sigma_r, \sigma_b)$, which essentially describes an assymetric gaussian absorption:
\begin{equation}
  \label{eq:fitlaursen}
  T(\lambda) = \left\{
    \begin{array}{ll}
      T_b - (T_b - T_{\rm min}) e^{-\frac{(\lambda-\lambda_0)^2}{2\sigma_b^2}} & \mathrm{if}\ \lambda < \lambda_0\\
      1 - (1 - T_{\rm min}) e^{-\frac{(\lambda-\lambda_0)^2}{2\sigma_r^2}} & \mathrm{if}\ \lambda \geq \lambda_0
    \end{array}\right.,
\end{equation}
where $\lambda_0$ is the central \lya wavelength, $T_{\rm min}$ is the minimum transmission, and $T_b$ correspond to the transmission far bluewards of \lya, scaled so that far from the line, the behaviour of $T(\lambda)$ follows closely the results from \citet{Laursen2011}, and $\sigma_r$ and $\sigma_b$ describe the width of the red and blue parts of the absorption line. We adjusted the parameters to get a correct rendering of the results of \citet{Laursen2011} at $z=2.5$ and $3.5$. We then interpolated each of the parameters to get the $z=3$ curve.

\begin{table}
\caption{Parameters for $T(\lambda)$}
\label{tab:laursen}
\centering
\begin{tabular}{lcccc}
  \hline\hline
  & $T_b$ & $T_{\rm min}$ & $\sigma_r$ & $\sigma_b$ \\
  \hline
  $z=2.5$ & 0.94 & 0.5 & 0.1 & 0.2\\
  $z= 3 $ & 0.87 & 0.3 & 0.125 & 0.275\\
  $z=3.5$ & 0.79 & 0.1 & 0.15 & 0.35\\
  \hline
\end{tabular}
\end{table}

\begin{figure*}
  \centering
  \resizebox{\hsize}{!}{\includegraphics{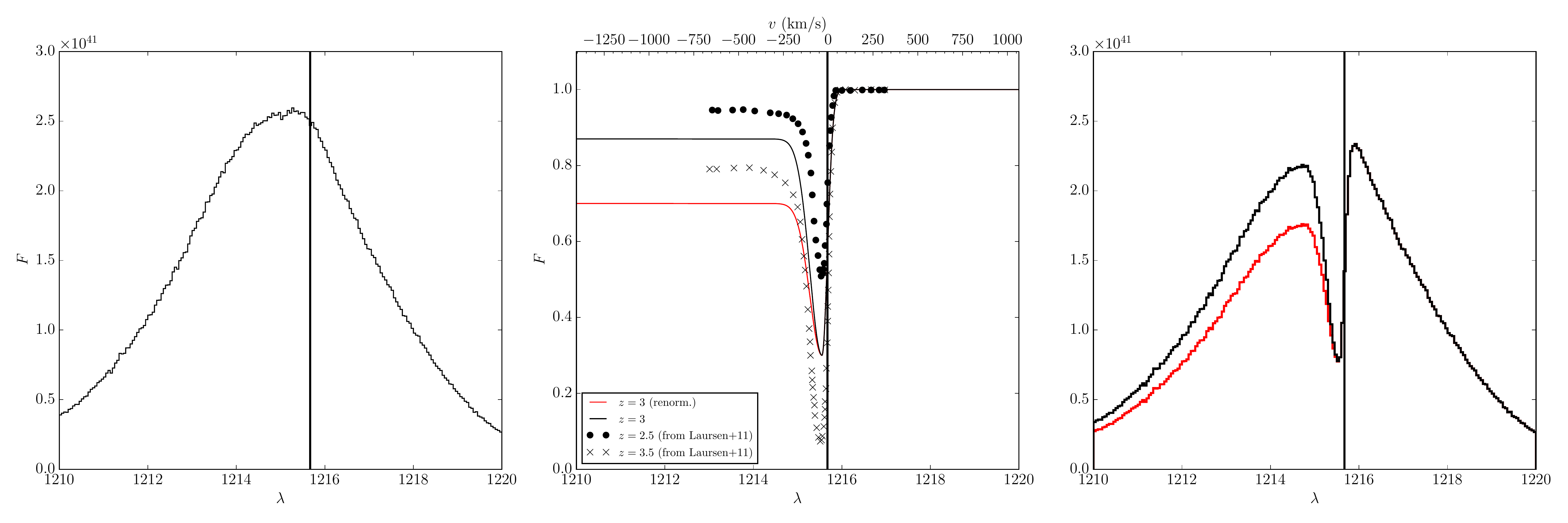}}
  \caption{\textit{Left}: Angle-averaged integrated spectrum of our blob (in black), divided in an extragalactic component (in red) and galactic emission (in blue).
    \textit{Middle}: IGM transmission function from \Rvir to the observer, fitting the results of \citeauthor{Laursen2011} at $z = 2.5$ (dotted line), $z=3$ (solid line) and $z=3.5$ (dashed line).
    \textit{Right}: Resulting spectrum after transmission through the IGM.
  For all three panels, the vertical line denotes the \lya wavelength.}
  \label{fig:laursen}
\end{figure*}

The left panel of Fig.~\ref{fig:laursen} illustrates the spectrum integrated over all directions of our blob as a black line, with the \lya wavelength indicated by a vertical line. We compute this spectrum right after the transfer inside the halo, so approximately at the virial radius. For the present experiment, we use the Gaussian model discussed in Sect.~\ref{sec:modelism} for the stellar component (see Fig.~\ref{fig:spectralshapes}). The other models yield the same results.
The central panel of Fig.~\ref{fig:laursen} presents the shape of the transmission function $T(\lambda)$ at $z = 3$ as a solid black line, and the data points extracted from \citet{Laursen2011} as circles ($z = 2.5$) and crosses ($z = 3.5$).
We give the parameters for our parametrization of $T(\lambda)$ in Table~\ref{tab:laursen}.
Using the method of \citet{Laursen2011}, we compute the observed spectrum as the multiplication of our spectrum at \Rvir with the IGM transmission function from \Rvir to the observer. The result of this is shown as the black line in the right-hand side panel of Fig.~\ref{fig:laursen}. We then integrate this to compute the transmitted fraction $\mathcal{T}_{\rm IGM}$, and conversely the fraction of all the photons that are scattered between \Rvir and the observer, $f$.
We find $\mathcal{T}_{\rm IGM} = 87\%$ or $f = 13\%$.

These values are relatively high compared to the canonical value of 0.67 for the transmission of the IGM at $z\sim 3$. This comes from the fact that the spectrum resulting from the transfer in our blob is very broad, because of the large velocity dispersion of the gas. This in turn means that the transmission is dominated by the very blue part of the spectrum. It is noteworthy that the \citet{Laursen2011} estimation of the transmission far from the \lya line gives much higher values than the observational estimates from e.g. \citet{Faucher-Giguere2008}. This overestimation is directly translated in an underestimation of $f$. To try to alleviate this issue and recover both the \lya forest constraints on the \lya transmission and the enhanced absorption in the vicinity of galaxies, we use an ad-hoc model for the transmission: using the same parametrisation as in Eq.~\ref{eq:fitlaursen}, we take the transmission blueward of \lya to be $T_b = 0.7$. The resulting transmission function is shown in red in the middle panel of Fig.~\ref{fig:laursen}, as is the transmitted spectrum in the right-hand side panel. For this model, we find $\mathcal{T}_{\rm IGM} = 78\%$, or alternatively $f = 22\%$, much closer to the $20\%$ inferred from the naive estimate presented in the text.

Using the results of \citet{Laursen2011}, we estimated the fraction of the luminosity transmitted from \Rvir to 5 \Rvir, which we will note $T(\Rvir, 5\Rvir)$, and approximate as $T(\Rvir,5 \Rvir) \sim T(\Rvir,\infty) / T(5 \Rvir, \infty)$. Here, $T(x,\infty)$ is the transmission between radius $x$ and the observer, which we have extracted from \citet{Laursen2011} as explained above. 
%Formally, the transmitted flux between two points A and B assuming that the medium in between is homogeneous can be expressed as $F_B = F_A \times T(A,B)$, where $T(A,B) \simeq e^{-\tau_{AB}}$. With this in mind, we can express $T(\Rvir, 5\Rvir) = \frac{T(\Rvir, \infty)}{T(5\Rvir, \infty)}$, where $T(x, \infty)$ is the transmission between a point $x$ and the observer.
We obtain $\mathcal{T}_{\Rvir, 5\Rvir} = 95\%$, meaning that only 5\% of the photons are scattered in a shell between \Rvir and 5 \Rvir. The assumption that 20\% of the luminosity is redistributed in a sphere of 5 \Rvir will therefore overestimate the impact of the IGM on the polarisation profile of our LAB. In the text, we show a model for which only 5\% of the photons are redistributed in that sphere of 5 \Rvir, but this time, it might very well underestimate the effect of the IGM. Indeed, if most of these photons have their locus of last scattering well inside the 5 \Rvir sphere, we should redistribute the luminosity in a much less wide area.
\begin{figure}
  \centering
  \resizebox{\hsize}{!}{\includegraphics{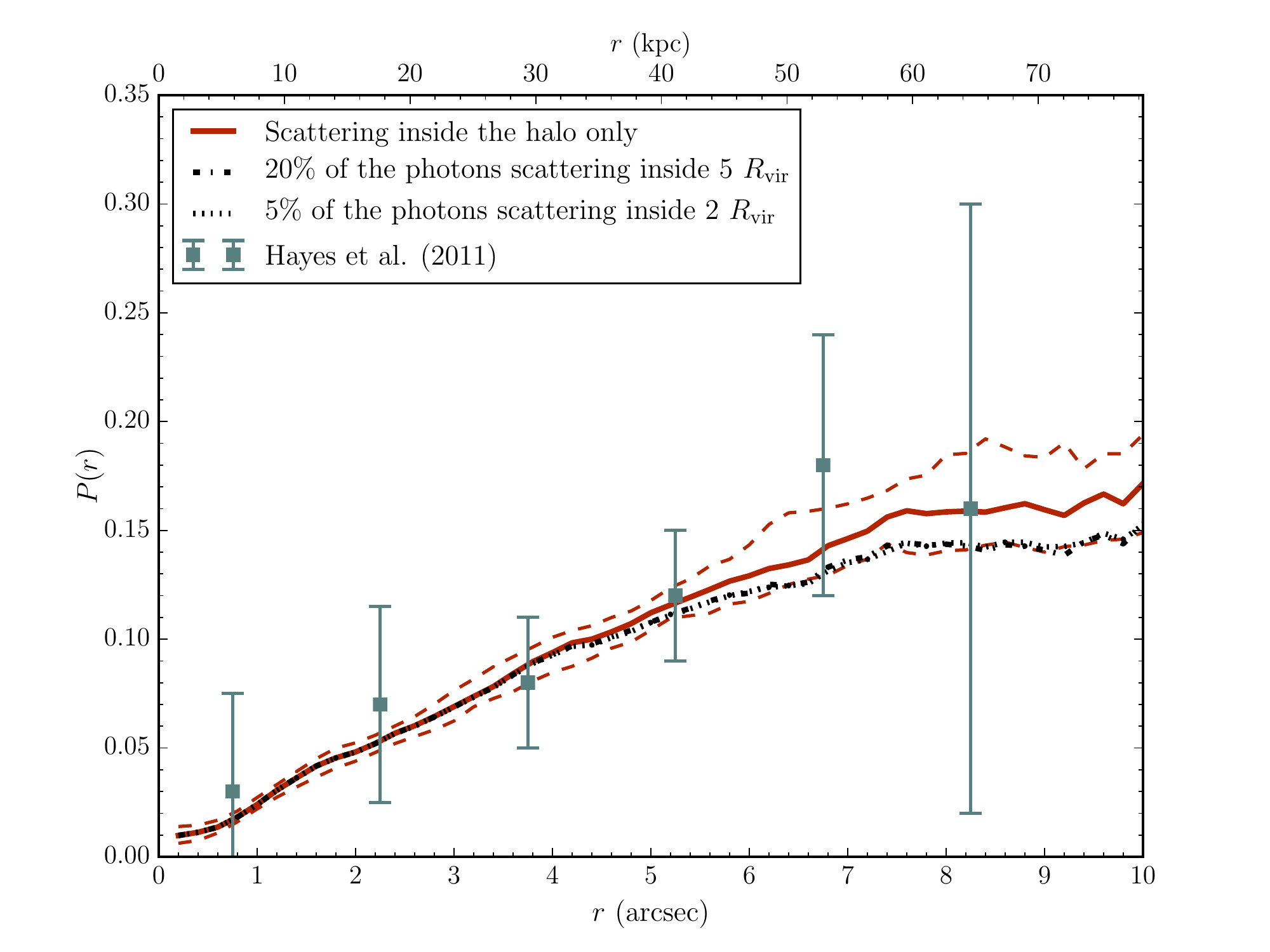}}
  \caption{Effect of the IGM on the polarization profile. We show the polarisation profile resulting from the redistribution of all the scattered photons in a sphere of 5 \Rvir as a dash-dotted line, and the one resulting from the redistribution of all the photons reaching 5 \Rvir in a smaller sphere of 2 \Rvir as a dotted line.}
  \label{fig:IGM_APP}
\end{figure}
On Fig~\ref{fig:IGM_APP}, we compare the impact of the scattering inside the IGM on the polarisation profile assuming either that 20\% of the luminosity is redistributed in a sphere of 5 \Rvir (dash-dotted black line) or that 5\% of the luminosity is redistributed in a sphere of 2 \Rvir (dotted black line). These two tentative overestimates of the effect of the IGM on our results produce similar results, which are indistinguishable from our raw prediction inside 40 kpc, and less than one standard deviation away at larger distances. Most of the photons that undergo scattering in the IGM are absorbed very far from the galaxy, so the luminosity must therefore be diluted in a very large area, and its impact on the polarisation is negligible.

In this appendix, we tested a more sophisticated method than in the main text to compute the fraction of photons which will scatter in the IGM beyond the viral radius, inspired from \citet{Laursen2011}.
It appears that this fraction of scattered photons is even smaller in this scenario.
To be conservative in our calculations, we assumed $f = 20\%$ in this work, consistent with an average transmission of $\mathcal{T}_{\rm IGM} \simeq 67\%$ for the blue part of the \lya line, and assuming that the red part of the line is left unchanged by the IGM.

\end{appendix}

\end{document}